\documentclass[acmsmall,screen,nonacm]{acmart}
\usepackage{placeins}
\usepackage{array,booktabs,longtable}
\AtBeginDocument{%
  }

\setcopyright{none}
\begin{document}

\title{Negotiating Future Visions with AI-Generated Images in Cooperative Design Workshops}

\author{Yuki Kubota}
\email{yuki.cs.kubota@ntt.com}
\correspondingauthor
\affiliation{%
  \institution{Communication Science Laboratories, NTT, Inc.}
  \city{Atsugi-shi}
  \state{Kanagawa}
  \country{Japan}
}
\author{Kazushi Maruya}
\email{kazushi.maruya@ntt.com}
\affiliation{%
  \institution{Communication Science Laboratories, NTT, Inc.}
  \city{Atsugi-shi}
  \state{Kanagawa}
  \country{Japan}
}

\author{Maki Ogawa}
\email{maki.ogawa@ntt.com}
\affiliation{%
  \institution{Communication Science Laboratories, NTT, Inc.}
  \city{Atsugi-shi}
  \state{Kanagawa}
  \country{Japan}
}

\author{Taiki Fukiage}
\email{taiki.fukiage@ntt.com}
\affiliation{%
  \institution{Communication Science Laboratories, NTT, Inc.}
  \city{Atsugi-shi}
  \state{Kanagawa}
  \country{Japan}
}

\author{Takumi Yokosaka}
\email{takumi.yokosaka@ntt.com}
\affiliation{%
  \institution{Communication Science Laboratories, NTT, Inc.}
  \city{Atsugi-shi}
  \state{Kanagawa}
  \country{Japan}
}

\author{Yuri Mikawa}
\email{yuri\_mikawa@ipc.i.u-tokyo.ac.jp}
\affiliation{%
  \institution{The University of Tokyo}
  \city{Tokyo}
  \country{Japan}}

\renewcommand{\shortauthors}{Kubota et al.}

\begin{abstract}
Generative AI is increasingly used to support collaborative creative practices, yet less is known about how AI-generated images reshape multiparty discussions through repeated cycles of revision. We examined a workshop in which participants envisioned future cities and technologies with AI-generated images. Drawing on audio-recorded discussions and other materials, we conducted a process-oriented qualitative analysis informed by the concepts of staging negotiation spaces and intermediary objects. Our analysis revealed four interrelated dynamics. (1) AI-generated images selectively materialized participants' concerns and sometimes introduced unexpected elements. (2) Participants interpreted and negotiated these images as provisional representations. (3) Through successive rounds of discussion, participants elaborated visual and spatial aspects of future cities, while purposes, situations of use, and infrastructural conditions received less sustained attention. Importantly, some concerns receded from subsequent discussions without being explicitly rejected, and such shifts were not always recognized as they evolved. (4) AI-generated images could also reshape participation by making some participants' contributions more central to subsequent exchanges. Our findings suggest that successive AI-generated images can reshape the negotiation space of cooperative design by making some concerns and participants' contributions more prominent than others, sometimes without explicit collective decisions. This highlights the importance of tracing how discussion topics and participation evolve across image revisions, and of developing AI tools and facilitation practices that make these shifts visible and help groups revisit concerns that have receded from discussion.
\end{abstract}

\ccsdesc[500]{Human-centered computing~Computer supported cooperative work}
\ccsdesc[500]{Human-centered computing~Empirical studies in collaborative and social computing}
\ccsdesc[300]{Human-centered computing~HCI design and evaluation methods}

\keywords{generative AI, cooperative design, qualitative analysis, intermediary objects, negotiation spaces}


\maketitle

\section{Introduction}
Generative AI has attracted considerable attention for its potential to transform creative practices traditionally undertaken by humans, including art and design~\cite{fu2020research,chiou2023designing,rafner2023creativity,heigl2026generative,wu2026human}. Recent studies have examined its effects on such creative practices, reporting improvements in creative performance under certain conditions and investigating how these effects vary across AI systems and visual stimuli~\cite{lee2024empirical,ge2025effects}. Beyond evaluating creative outcomes, research on image-generative AI has explored how people use generated images to formulate problems, explore design possibilities, and develop novel ideas~\cite{chiou2023designing,rafner2023picture}.

Generative AI is also being explored in cooperative work, including collaborative writing, team-based problem solving, and communication between designers and users~\cite{michel2025collaborative,wei2025effects,he2024revealing}. In cooperative design, visual materials have long served as resources for communicating ideas and discussing design alternatives~\cite{shen2010online,akbar2025use}. Although producing detailed visualizations can require considerable time and specialized skills, generative AI enables images to be generated and revised through textual instructions, potentially making visual exploration more accessible~\cite{verheijden2023collaborative,von2024equipping}. Recent studies have explored its use in participatory public-space and landscape design~\cite{guridi2025image,awashra2025generative} and in envisioning future cities~\cite{von2024equipping}, as well as related applications involving architectural ideation, cultural heritage, post-conflict urban landscapes, and future robotic technologies~\cite{paananen2024using,he2025recall,alkhateeb2024post,hoggenmueller2023creative}.

However, the role of visual materials, including AI-generated images, extends beyond  communicating ideas. Research on intermediary objects in collaborative design has shown that sketches, drawings, and digital models can mediate collaborative work by enabling participants to interpret and develop design ideas together~\cite{boujut2003intermediary,paavola2019dynamics,vinck2012accessing}. At the same time, a longstanding concern in design research is design fixation, in which existing examples or solutions limit the exploration of alternatives~\cite{youmans2014design,crilly2017next}. Studies of AI-supported ideation have reported both benefits and limitations in this regard. AI-generated images have been rated as more inspiring, enjoyable, and useful than conventional image search~\cite{cai2023designaid}, while an experimental study found that participants using AI-generated images exhibited greater design fixation and produced fewer original ideas~\cite{wadinambiarachchi2024effects}. These findings highlight the importance of examining how AI-generated images shape collaborative design processes.

Despite these insights, it remains insufficiently understood how AI-generated images become involved in the development of multiparty discussions. Research on collaborative image generation identified the ``crooked bow-tie effect,'' describing difficulties in translating complex design intentions into textual prompts and corresponding images~\cite{verheijden2023collaborative}. Workshops with design students showed that text-to-image AI encouraged verbal articulation and supported semantic convergence as participants shared interpretations of generated images~\cite{lin2024text}. Research on participatory public-space design further showed that imperfect AI-generated images could stimulate conversations about public spaces~\cite{guridi2024fake}. Nevertheless, less attention has been paid to how AI-generated images shape collaborative design through repeated cycles of discussion and image generation. In particular, it remains unclear how participants' ideas are represented in successive images, how those images are interpreted and revised, and how these interactions shape subsequent discussions and participation.

To investigate these processes, we examined a workshop using AI-generated images to support collaborative discussion. Eight groups envisioned future cities and potential applications of digital-signage technologies through successive rounds of discussion and image generation. Using audio recordings and other workshop materials, we traced how discussion topics and participation evolved through successive image revisions. Our process-oriented qualitative analysis was informed by two related theoretical perspectives: staging negotiation spaces~\cite{pedersen2020staging} and intermediary objects~\cite{vinck2012accessing,sadorge2024generative}. Through this analysis, we address the following central research question: How does the negotiation space of a collaborative design workshop evolve through repeated cycles of image generation and discussion?

\section{Theoretical backgrounds}
We draw on two related theoretical perspectives that serve distinct roles in this study. The framework of staging negotiation spaces~\cite{pedersen2020staging,pedersen2022navigating,pedersen2025staging} guides our process-oriented examination of how participants' concerns and interactions evolve across workshop discussions, whereas the concept of intermediary objects~\cite{vinck1994mediating,vinck2012accessing} situates AI-generated images within broader research on how design artifacts mediate collaborative work.

\subsection{Staging negotiation spaces}
Staging negotiation spaces is an analytical framework for understanding collaborative design processes in which designers engage with multiple concerns held by various involved actors~\cite{pedersen2020staging,pedersen2022navigating,pedersen2025staging}. The framework expands the role of designers from creative experts to directors and facilitators of negotiations, using the metaphor of an inviting stage director to emphasize how designers configure the conditions for collaborative engagement. From this perspective, a design or research process can be understood as a series of negotiation spaces. Staging such spaces involves inviting relevant actors to participate and articulate salient concerns, developing design-related objects through which negotiation can take place, and aligning these objects with the material arrangement of the setting. The framework further emphasizes the iterative character of this process, in which negotiations are interpreted and reframed across successive rounds of staging.

The theoretical background of this framework lies in actor-network theory (ANT) and its subsequent uptake in design research~\cite{latour2005reassembling,storni2015designing}. Rather than explaining social phenomena through pre-given structures, institutions, or social groups, ANT examines how associations among heterogeneous human and non-human actors are formed, maintained, and stabilized. In design contexts, non-human actors may include sketches, models, photographs, and digital technologies, which can shape how problems are framed through their interactions with people. This perspective has informed various strands of design research that conceptualize participation and design as socio-material processes, including temporary spaces of innovation~\cite{clausen2015social}, design multiples~\cite{fischer2021design}, participatory infrastructuring~\cite{bodker2017tying}, and design things or object-oriented publics~\cite{ehn2011participation,jenkins2016object}. Across these perspectives, design outcomes and possibilities are understood as emerging from interactions among participants and the material resources.

Building on this perspective, we approached the workshop as an evolving negotiation space involving participants, facilitators, AI-generated images, and other resources. We drew on this framework to structure our group case notes and cross-case comparisons, examining how concerns, interactions, and the roles of different actors and resources evolved throughout the workshop.

\subsection{Intermediary objects}
The concept of intermediary objects focuses on intermediate products that are produced and transformed within collaborative work, originating from an ANT-inspired view~\cite{vinck1994mediating,vinck2012accessing}. It draws attention to how such artifacts help organize collaborative activities as design processes evolve. In collaborative design, such objects can  facilitate shared understanding and the collective development of design ideas~\cite{boujut2003intermediary}. Importantly, they may also carry inscriptions of values, assumptions, and constraints that shape their subsequent discussion~\cite{akrich1992description}.

Intermediary objects have been used as an analytical framework in studies of collaborative design and computer-supported cooperative work, including architectural and engineering settings~\cite{eckert2003role,paavola2019dynamics,wither2025using}. More recently, scholars have examined the generative or performative roles of such objects~\cite{wither2025using,sadorge2024generative}. A study of healthcare infrastructure design illustrates how intermediary objects shape the exploration of design problems and possible solutions~\cite{sadorge2024generative}. This study identifies two dynamics: ``branching out,'' in which intermediary objects broaden exploration by introducing new problem framings and possibilities, and ``unfolding,'' in which they deepen exploration by elaborating existing problems and possible solutions.

Some previous research on AI-generated images in collaborative design provides a basis for considering their relationship to intermediary objects. Studies have identified difficulties in translating design intentions into generated images~\cite{verheijden2023collaborative}, while also showing that imperfect images can serve as resources for discussion~\cite{guridi2024fake}. In this study, we treat AI-generated images as intermediary objects, connecting our findings to broader research on design artifacts in collaborative work. This perspective informs our discussion of both idea development and shifts in attention and participation across successive image revisions.

\section{Data and methods}
This section describes the workshop context and procedures, the materials used, and the methods of data collection and analysis. The study was approved by the Research Ethics Committee of our institution and conducted in accordance with the Declaration of Helsinki. Written informed consent was obtained from all participants. 

In preparing this manuscript, the authors used ChatGPT (OpenAI) solely to assist with English-language editing. All AI-assisted revisions were carefully reviewed by the authors, who take full responsibility for the final manuscript.

\subsection{Study context and participants}

\begin{table}[t]
\centering
\caption{Participant demographics, interests in science and technology, and prior experience with generative AI.}
\label{tab:participants}
\small
\begin{tabular}{llllll}
\toprule
\textbf{ID} &
\textbf{Group} &
\textbf{Age} &
\textbf{Gender} &
\textbf{S\&T interest} &
\textbf{Prior GenAI use} \\
\midrule
P01 & 9/6-A & 60 or older & Male & Very interested & Every 1--2 weeks \\
P02 & 9/6-A & 50s & Female & Very interested & Several times/day or more \\
P03 & 9/6-A & 40s & Female & Very interested & Never \\
P04 & 9/6-A & 40s & Female & Interested & Every 1--2 weeks \\
P05 & 9/6-A & 40s & Male & Interested & Every 1--2 weeks \\
\midrule
P06 & 9/6-B & Elementary school & Female & Interested & Several times/day or more \\
P07 & 9/6-B & Elementary school & Female & Neutral & Several times/day or more \\
P08 & 9/6-B & Elementary school & Male & Very interested & A few times \\
\midrule
P09 & 9/6-C & 40s & Male & Interested & A few times \\
P10 & 9/6-C & 50s & Female & Very interested & Several times/day or more \\
P11 & 9/6-C & 60 or older & Female & Very interested & Never \\
P12 & 9/6-C & 60 or older & Male & Very interested & Every 1--2 days \\
\midrule
P13 & 9/6-D & 20s & Female & Interested & Every 1--2 days \\
P14 & 9/6-D & 40s & Male & Very interested & Every 1--2 months \\
P15 & 9/6-D & 40s & Male & Interested & Several times/day or more \\
P16 & 9/6-D & 20s & Male & Not interested & Never \\
\midrule
P17 & 9/7-A & 40s & Female & Neutral & Never \\
P18 & 9/7-A & 40s & Female & Interested & A few times \\
P19 & 9/7-A & 40s & Male & Very interested & Several times/day or more \\
P20 & 9/7-A & 50s & Female & Interested & Every 1--2 weeks \\
\midrule
P21 & 9/7-B & Elementary school & Male & Interested & Every 1--2 days \\
P22 & 9/7-B & Elementary school & Male & Not sure & Never \\
P23 & 9/7-B & Elementary school & Male & Very interested & Every 1--2 days \\
P24 & 9/7-B & Elementary school & Male & Interested & A few times \\
\midrule
P25 & 9/7-C & 20s & Male & Very interested & Several times/day or more \\
P26 & 9/7-C & 30s & Male & Very interested & Never \\
P27 & 9/7-C & 20s & Male & Interested & A few times \\
\midrule
P28 & 9/7-D & 20s & Female & Interested & Several times/day or more \\
P29 & 9/7-D & 20s & Male & Very interested & Several times/day or more \\
P30 & 9/7-D & 20s & Female & Interested & Every 1--2 days \\
P31 & 9/7-D & 20s & Male & Very interested & Every 1--2 days \\
\bottomrule
\end{tabular}
\end{table}

\begin{figure}[htb]
  \centering
  \includegraphics[keepaspectratio,width=1.00\linewidth]{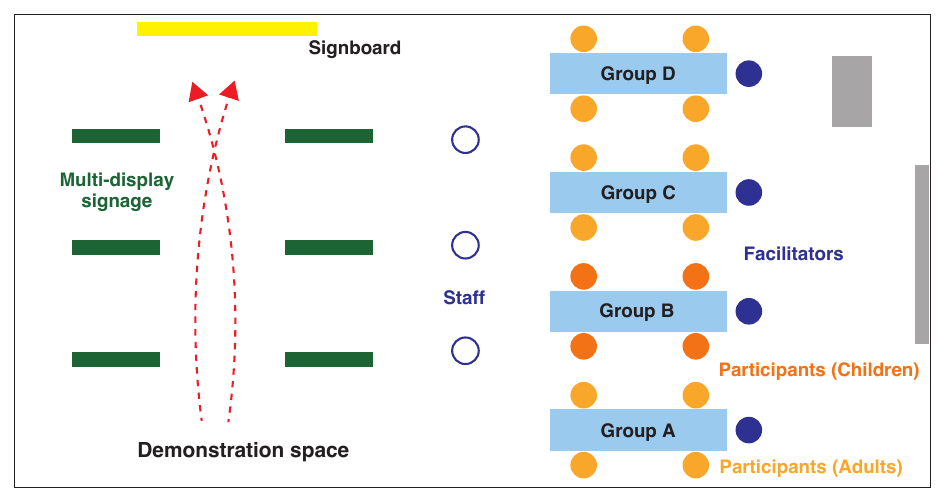}
  \caption{Spatial arrangement during the workshop. Participants first experienced demonstrations of multi-display signage (left) and then took part in the workshop in four separate groups (right).}
  \label{fig:workshop-layout} 
\end{figure}

The study examined a workshop held at a science museum in Japan. The workshop focused on potential applications of multi-display digital signage, a technology examined in prior research on pedestrian behavior~\cite{mikawa2026visual} and public perceptions of digital nudging~\cite{kubota2026nudging}. Participants were invited to envision future urban environments and everyday life. 

Two 90-minute workshop sessions were held on September 6 and 7, 2025. Figure~\ref{fig:workshop-layout} illustrates the spatial arrangement of the workshop. Each session comprised demonstrations of multi-display signage and two group activities: Work 1, in which participants compared digital signage with familiar information and communication technologies, and Work 2, in which they collaboratively imagined future urban spaces and technologies using AI-generated images. The present study focuses on the discussions conducted during Work 2.

Participants were recruited through an announcement on the science museum's website, with interested individuals registering in advance. Table~\ref{tab:participants} summarizes the demographic and background characteristics of the participants. A total of 31 participants attended the workshops, with 16 participating on the first day and 15 on the second. Participants ranged from elementary-school-aged children to adults aged 60 or older, with varying levels of experience with generative AI. Most participants (87.1\%) reported an interest in science and technology. Nearly half (48.4\%) reported using generative AI every one or two days or more frequently, whereas 19.4\% reported never having used it.

Participants were divided into four groups on each day, yielding eight groups of three to five members. Groups were formed with attention to age composition and to encourage interaction among previously unacquainted participants. On both days, Group B consisted primarily of elementary-school-aged children, whereas the other groups comprised mostly university-age or older participants. Four authors served as group facilitators, while one additional author and two science museum staff members supported the activities and occasionally joined the discussions. All the authors were researchers in information technology or psychology involved in digital signage research and familiar with AI and digital signage technologies. The museum staff were non-research personnel.

\subsection{Workshop procedure}
\begin{table}[htb]
    \centering
    \caption{Overview of the workshop procedure. The present study focuses on Work 2, in which participants collaboratively envisioned future urban spaces and explored possible applications of digital signage using AI-generated images.}
    \label{tab:procedures-whole}
    \begin{tabular}{p{2.5cm}p{4.5cm}p{6.5cm}}
        \hline
        \textbf{Time} & \textbf{Items} & \textbf{Contents} \\ \hline
        00:00--00:05 & Introduction &
        The overall theme of the event was introduced. \\
        00:05--00:40 & Demonstrations &
        Participants experienced multi-display digital signage through two demonstrations. \\
        00:40--00:55 & Work 1: Discussing characteristics of digital signage technology &
        Participants compared digital signage with familiar information and communication technologies. \\
        00:55--01:25 & Work 2: Envisioning future cities with AI-generated images &
        Participants explored possible uses of digital signage and envisioned everyday life in a future city.\\
        01:25--01:30 & Answering questionnaire &
        Participants completed the post-workshop questionnaire.\\ \hline
    \end{tabular}
\end{table}

Table~\ref{tab:procedures-whole} summarizes the workshop procedure and the time allocated to each activity. Before the group activities, participants were introduced to two basic guidelines, which had also been shared with the group facilitators. First, they were encouraged to listen to others and avoid dominating the discussion so that everyone had an opportunity to contribute. Second, they were invited to share tentative ideas and explore future cities and technologies from multiple perspectives. The following subsections describe the activities, with particular emphasis on Work 2, which formed the primary focus of our analysis.

\subsubsection{Demonstrations}
Participants first experienced two demonstrations using multiple displays installed at the rear of the workshop room (Figure~\ref{fig:workshop-layout}). In the first demonstration, moving stripe patterns were presented to induce vection and examine whether visual motion could influence participants’ walking trajectories. In the second, immersive videos, including seaside and outer-space scenes, were presented to elicit affective experiences. Although the specific video conditions differed slightly between the two workshop days, both sessions included these two types of demonstrations.

\subsubsection{Work 1: Discussing characteristics of digital signage technology}
Work 1 served as a preparatory activity for Work 2, encouraging participants to consider the characteristics and possible uses of digital signage in relation to familiar information and communication technologies. Participants selected a technology from the handout (Appendix~\ref{app:work1}, Figure~\ref{fig:procedures-ws1}) and identified similarities and differences between it and digital signage on sticky notes. They then exchanged the sheets with their sticky notes attached within their groups to add further observations. This activity provided an opportunity to consider possible applications of digital signage before envisioning its roles in future cities. Further details are provided in Appendix~\ref{app:work1}.

\subsubsection{Work 2: Envisioning future cities with AI-generated images}
\begin{table}[htb]
    \centering
    \caption{Detailed procedure of Work 2.}
    \label{tab:procedures-work2}
    \begin{tabular}{p{5cm}p{9cm}}
        \hline
        \textbf{Items} & \textbf{Contents} \\ \hline
        (1) Introduction &
        The facilitator explained the purposes and procedure of Work 2. \\
        (2) Selection of a place &
        Each group selected a shared setting for envisioning a future city. \\
        (3) Individual reflection using a worksheet &
        Participants wrote their individual ideas about the future city using the worksheet. \\
        (4) Iterative discussion and image generation (2--4 rounds) &
        Participants discussed their ideas, while facilitators generated updated images based on the discussion over multiple rounds. \\
        (5) Final presentation &
        A representative from each group presented its final AI-generated image and shared the main ideas developed through discussion with all workshop participants. \\
        \hline
    \end{tabular}
\end{table}
\begin{figure*}[htb]
  \centering
  \includegraphics[keepaspectratio,width=1.00\linewidth]{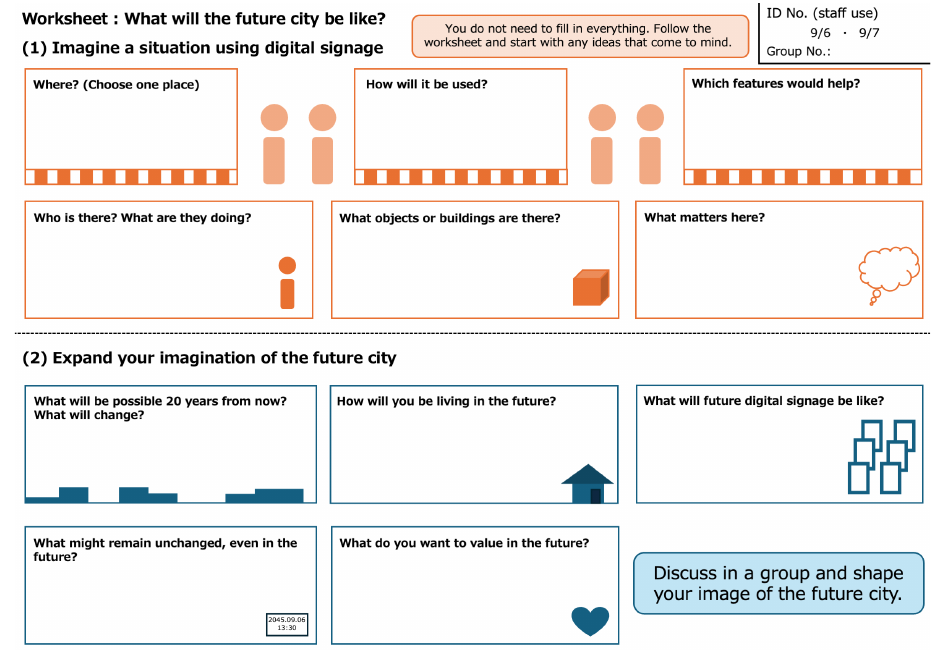}
  \caption{English translation of the worksheet used for individual reflection in Work 2. The orange section instructed participants to consider potential uses of digital signage in future urban settings, and the blue section encouraged broader reflection on future everyday life and values.}
  \label{fig:procedures-ws2}
\end{figure*}
\begin{figure*}[htbp]
  \centering
  \includegraphics[keepaspectratio,width=0.90\linewidth]{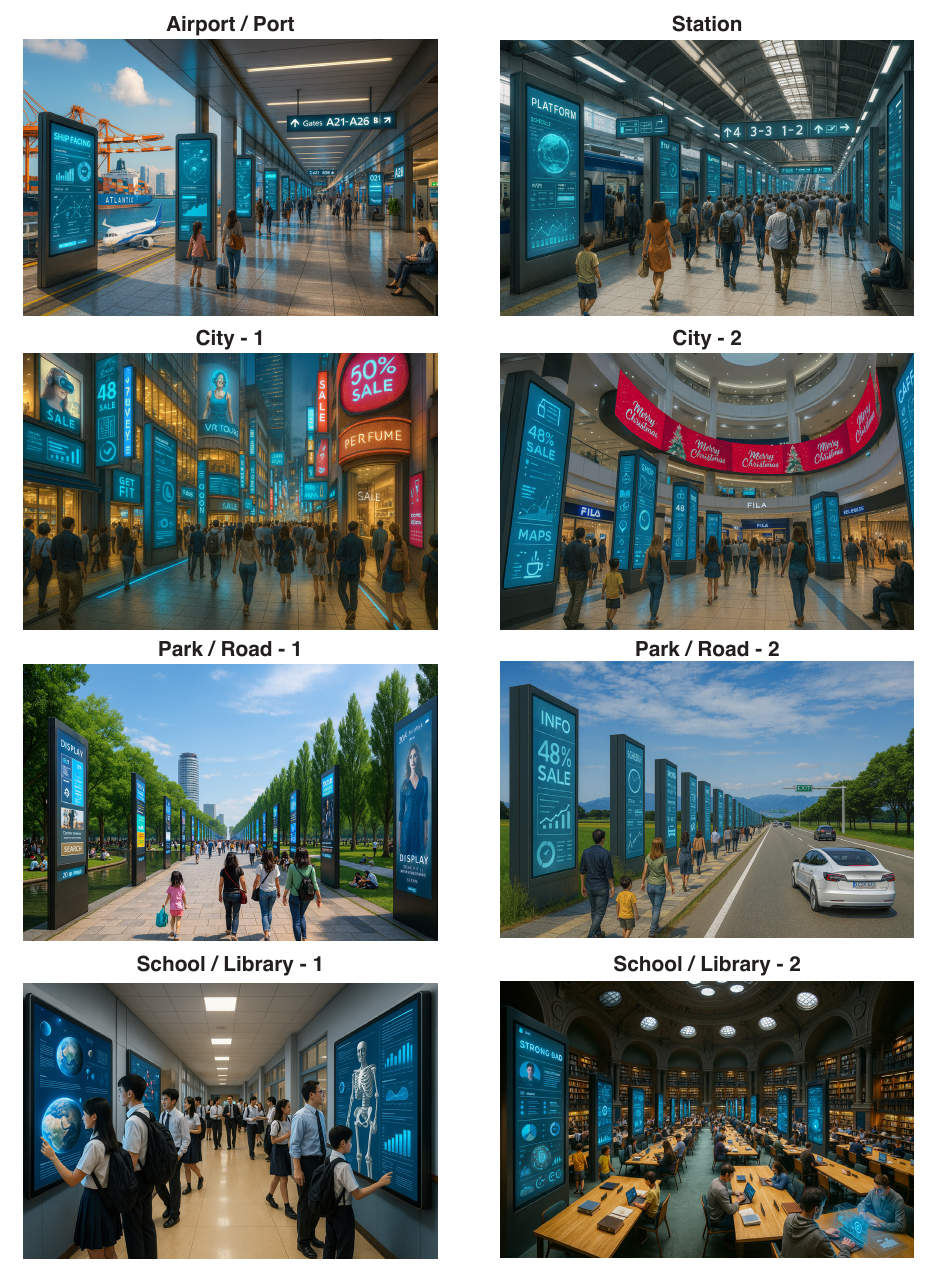}
  \caption{Initial AI-generated images prepared for Work 2. Facilitators prepared one or two images in advance for each of the five urban-setting categories using the image-generation functionality of ChatGPT (GPT-5). The images depicted future urban environments incorporating digital signage.}
  \label{fig:base-generated-images}
\end{figure*}
\begin{figure*}[htb]
  \centering
  \includegraphics[keepaspectratio,width=1.00\linewidth]{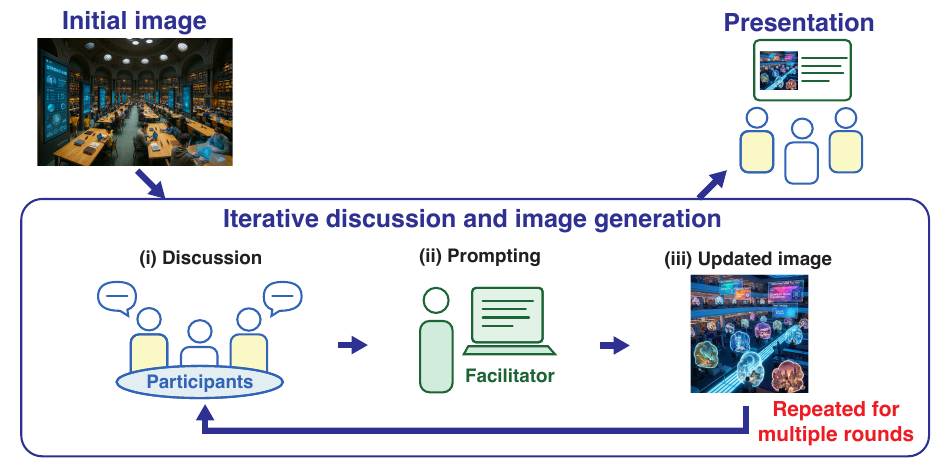}
  \caption{Iterative discussion and image generation in Work 2. Starting from an initial AI-generated image, participants discussed their ideas, which the facilitator translated into prompts for image generation. The updated images were presented for further discussion, and this cycle was repeated over multiple rounds (Table \ref{tab:procedures-work2}, Item 4). Finally, each group presented its final image and main ideas to all participants (Table \ref{tab:procedures-work2}, Item 5).}
  \label{fig:procedures-discussion}
\end{figure*}

Work 2 focused on collaboratively envisioning future urban environments and exploring possible applications of digital signage (Table~\ref{tab:procedures-work2}). Participants were first presented with photographs representing five categories of urban settings: airport/port, station, city, park/road, and school/library. Each participant selected a preferred setting, and when preferences differed within a group, group members agreed on a setting to explore together. This selection established a shared spatial context for subsequent discussion and image generation.

Participants then individually considered their ideas about the selected setting using a worksheet (Figure~\ref{fig:procedures-ws2}). The worksheet comprised two sections: the first (orange) instructed participants to consider concrete situations in which digital signage might be used, while the second (blue) encouraged broader reflection on future everyday life and values. Participants were encouraged to start with whichever items they found easiest and write down tentative ideas, but were not required to complete every item.

After approximately five minutes of individual reflection, facilitators presented each group with an initial AI-generated image of its selected setting (Figure~\ref{fig:base-generated-images}). These images had been prepared in advance and did not incorporate participants’ worksheet responses. Participants then used these images as a starting point for group discussion.

As illustrated in Figure~\ref{fig:procedures-discussion}, the collaborative work followed a repeated sequence of (i) discussion, (ii) prompting, and (iii) presentation of an updated image. In each round, participants shared their ideas and discussed them with the current image. The first round used the pre-generated initial image (Figure~\ref{fig:base-generated-images}), which did not incorporate participants' ideas; subsequent rounds used images generated from the preceding discussion. Facilitators then translated the discussion into image-generation prompts. Facilitators primarily used Gemini 2.5 Flash Image, with the image-generation functionality of ChatGPT (GPT-5) also used in some groups. They could generate multiple candidates before selecting an image to present to the group. The updated image then served as the starting point for the next round of discussion, allowing participants to reconsider and develop their ideas further. This cycle was repeated for two to four rounds. Facilitation practices varied across groups, and other staff members occasionally joined the discussions, particularly to comment on technical feasibility or possible future developments.

At the end of Work 2, facilitators prepared summary slides containing each group's final AI-generated image and brief notes on the ideas discussed. A representative from each group then explained the image and main ideas to all workshop participants. During the final presentations, a moderator (one of the facilitators) facilitated cross-group sharing and occasionally asked follow-up questions to clarify the groups' proposals. Following the workshop, participants completed the post-workshop questionnaire described in Appendix~\ref{app:questionnaire}.

\subsection{Data collection and analysis}
\label{sec:data-collection}
We collected several types of data during and after the workshop, including audio recordings and transcripts of the Work 2 discussions, participants’ worksheets, facilitators' image-generation prompts, AI-generated images presented to the groups, and post-workshop questionnaire responses. The questionnaire responses were analyzed quantitatively to characterize participants' backgrounds and examine their retrospective perceptions of the workshop activities and use of generative AI (Appendices~\ref{app:questionnaire} and \ref{app:questionnaire-results}). 

For the qualitative analysis, we drew on reflexive thematic analysis~\cite{braun2019reflecting,braun2022thematic} to examine how participants' ideas and discussions developed during Work 2. Discussion transcripts served as the primary data source, supplemented by other workshop materials. The first author led the qualitative analysis, combining iterative coding and theme construction with a process-oriented examination of each group's discussions. We iteratively revisited the source materials, codes, and developing themes to refine our interpretations, while reflecting on our involvement in designing and facilitating the workshop (see Section~\ref{sec:reflexivity}).

Following familiarization with each group's discussions, the first author coded all textual materials in at least three passes. The first pass involved open coding to develop provisional codes, which were subsequently consolidated and provisionally defined. During the second and third passes, the first author refined code labels, definitions, and boundaries before establishing the codes used for case analysis. The developing code labels and definitions were also discussed with other authors. Coding addressed three aspects: content (participants' concerns and proposed applications), structure (workshop stages and image-revision cycles), and performance (the articulation of future visions, interpretation of AI-generated images, and negotiation of ideas and participation). These codes served as analytical resources for organizing and revisiting the materials. The resulting codes and their definitions are provided in Appendix~\ref{app:coding-system} (Tables~\ref{tab:codebook-structure}--\ref{tab:codebook-performance}).

The first author then conducted a process-oriented analysis of each group using a nine-section case-note template developed with reference to the framework of staging negotiation spaces. The first three sections followed the workshop sequence: (1) initial concerns and values; (2) discussions during image generation and after image presentation; and (3) final group presentations. The remaining sections examined aspects spanning these stages: (4) changes in topics and concerns; (5) relationships between worksheets and group discussions; (6) roles of facilitators and staff; (7) representation and interpretation of AI-generated images; (8) negotiation and articulation of future visions; and (9) other relevant observations.

For the cross-case analysis, the first author synthesized the eight case notes around three aspects: (i) shifts in discussion focus, (ii) the roles of participants, facilitators, AI-generated images, and other resources, and (iii) the overall structure and dynamics of discussion. The first author recorded key observations on individual cards and iteratively compared and grouped them to identify recurring patterns and differences across groups. Through this process, the first author developed interpretive themes concerning the roles of AI-generated images in collaborative discussions.

Other authors reviewed the coding system, individual case notes, and findings from the cross-case analysis, with those who served as facilitators focusing particularly on the groups they had facilitated. Differences in interpretation were discussed among the authors in relation to the workshop records, contributing to the refinement of the analysis and themes. The case-level accounts and cross-case overview are provided in Appendix~\ref{app:workshop-trajectories}.

\subsection{Researcher positionality and reflexivity}
\label{sec:reflexivity}
Several authors were involved in designing and facilitating the workshop. This involvement provided firsthand knowledge of the workshop's objectives, procedures, and underlying technologies, but also introduced the possibility that the analysis might overemphasize the intended effects of the workshop design or the role of generative AI. Moreover, by translating participants' ideas into prompts and selecting images for presentation, the facilitators were themselves involved in shaping the interactions examined in this study. During the workshop, the facilitators sought to encourage participants to articulate and develop their own perspectives without steering discussions toward predetermined conclusions.

The first author, who led the qualitative analysis, approached the study with a generally positive view of the potential of generative AI in participatory workshops. During facilitation, the first author also formed an initial impression that AI-generated images might draw participants' attention toward visually representable aspects of future cities, potentially narrowing the range of topics discussed. To critically examine this initial impression, the first author conducted the analysis in successive stages, from broad coding to group case notes and cross-case comparisons. At each stage, the first author revisited the original materials and considered alternative interpretations, including cases in which participants' concerns persisted or were rearticulated through different proposals, cases in which AI-generated images appeared to have limited influence, and cases in which other actors and resources contributed to shifts in discussion.

Discussions among the authors also formed part of this reflexive process. The authors brought different perspectives to the workshop episodes and developing interpretations, drawing on their experiences of designing or facilitating the workshop, supporting its activities, or examining the workshop records. These exchanges provided opportunities to critically examine the first author's interpretations and consider alternative explanations.

\section{Results}
\label{sec:results}
\begin{figure}[htb]
  \centering
  \includegraphics[keepaspectratio,width=1.0\linewidth]{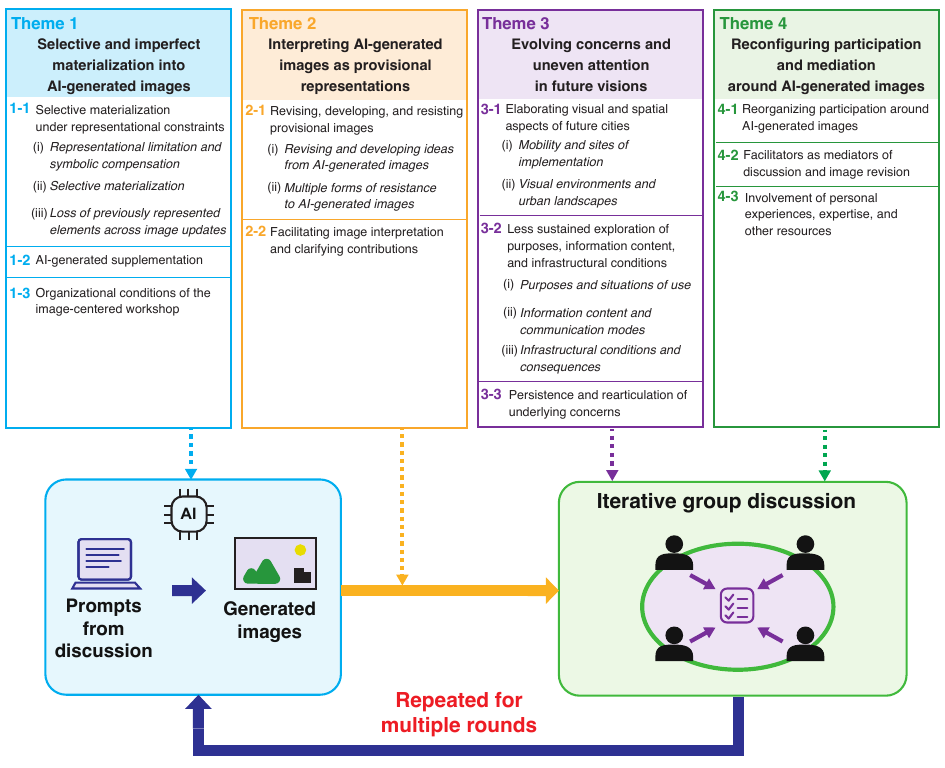}
  \caption{Overview of the four themes and their sub-themes identified in this study. The themes represent interconnected aspects of the iterative process of AI-image generation and collaborative discussion. Solid arrows indicate the workshop process, while dotted lines indicate the primary analytical focus of each theme.}
  \label{fig:results-themes}
\end{figure}

Our analysis identified four interrelated themes concerning how AI-generated images became involved in and shaped collaborative discussions (Figure~\ref{fig:results-themes}). These themes capture distinct but interconnected aspects of the iterative process of image generation and group discussion:

\begin{enumerate}
    \item \textbf{Theme 1}---Selective and imperfect materialization into AI-generated images
    \item \textbf{Theme 2}---Interpreting AI-generated images as provisional representations
    \item \textbf{Theme 3}---Evolving concerns and uneven attention in future visions
    \item \textbf{Theme 4}---Reconfiguring participation and mediation around AI-generated images
\end{enumerate}

The following sections elaborate on each theme through representative episodes, tracing how participants' discussions developed around the generated images. The original discussions were conducted in Japanese. Excerpts presented in English were translated and, where indicated, abbreviated for readability. In the excerpts, participant labels combine the group letter and participant number (e.g., PC-1), with numbering defined separately for each group on each workshop day. These speaker labels could not be reliably matched to the participant IDs in Table~\ref{tab:participants} using the audio recordings alone. The label F denotes the facilitator, and groups are identified by date and letter (e.g., 9/6-C).

\subsection{Theme 1: Selective and imperfect materialization into AI-generated images}
This theme captures how participants' concerns were selectively materialized into AI-generated images through successive rounds of discussion, prompting, and image generation. While some ideas  became visually prominent, others were only partially represented or omitted, and the generated images sometimes introduced elements that participants had not explicitly proposed. Prior research has identified mismatches between design intentions and generated images (e.g., the ``crooked bow-tie effect''~\cite{verheijden2023collaborative}) and highlighted how imperfect images can stimulate collaborative discussion~\cite{guridi2024fake}. We examine these transformations in relation to selective materialization under representational constraints (Theme 1-1), AI-generated supplementation (Theme 1-2), and the conditions under which image generation was organized in the workshop (Theme 1-3).

\subsubsection{Theme 1-1: Selective materialization under representational constraints}
\paragraph{(i) Representational limitation and symbolic compensation}
\begin{figure}[htb]
  \centering
  \includegraphics[keepaspectratio,width=1.0\linewidth]{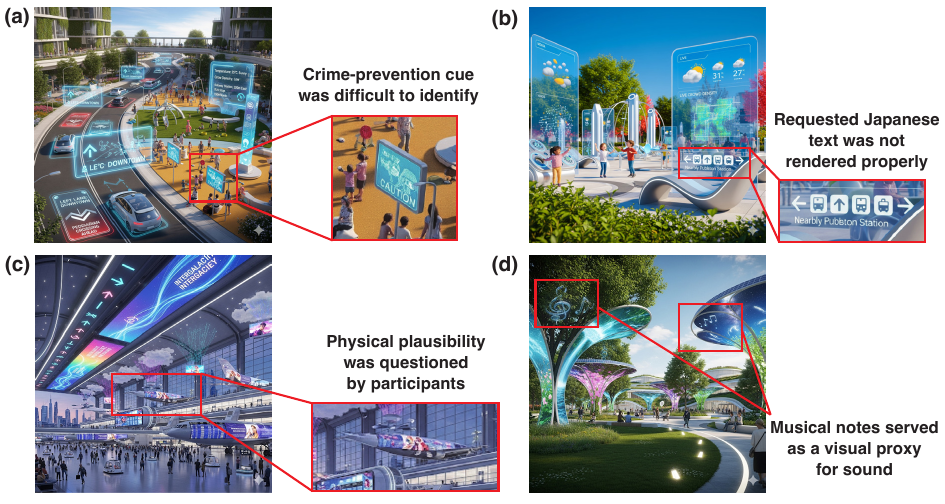}
  \caption{Theme 1-1(i): Examples of representational limitations and symbolic compensation in AI-generated images. (a) In Group 9/6-C, a crime-prevention cue was difficult to identify. (b) In Group 9/6-C, Japanese text was not adequately rendered despite revised prompting. (c) In Group 9/7-B, participants questioned the plausibility of implied movement. (d) In Group 9/7-A, musical notes served as a visual proxy for sound. Red boxes highlight the relevant image elements, with enlarged details shown where appropriate.}
  \label{fig:results-1-1-limitation}
\end{figure}

The static images used in the workshop could not directly convey nonvisual or dynamic aspects of participants' ideas, such as sound and motion. The generative AI also sometimes struggled to render written text accurately, particularly Japanese. These representational and technical constraints meant that some aspects of participants' envisioned futures could not be adequately materialized in the generated images.

For example, in Group 9/6-C, participants initially proposed ideas related to safety warnings and everyday information provision in streets and parks. After viewing the generated image (Figure~\ref{fig:results-1-1-limitation}a), they questioned whether several aspects of their discussion had been represented:

\begin{itemize}
\item[PC-1:] \textit{So, can the AI only produce images? What about sound?}
\item[F:] \textit{What we are using here only generates static images. There are various types of AI, and some can generate audio as well.}
\item[PC-2:] \textit{Is the idea of crime prevention that I mentioned represented anywhere in this image?}
\item[F:] \textit{It is extremely small. Around here [a sign near the lower-right area of the image], there is a sign that says ``Caution'' in English, so perhaps that is what AI represented.}
\item[PC-3:] \textit{I am wondering if it can be written in Japanese. Is that difficult?}
\item[F:] \textit{It depends somewhat on the model, but I can try asking it to generate the text in Japanese.}
\end{itemize}

This exchange highlights difficulties in representing the intended forms of information provision, including auditory information and Japanese-language text. Although the facilitator identified a small sign as a possible representation of the initial idea, its connection to the participant's proposal was not immediately apparent. The facilitator subsequently revised the prompt to request Japanese text explicitly, but the updated image (Figure~\ref{fig:results-1-1-limitation}b) still failed to render it adequately. The facilitator then remarked that the AI \textit{``seems to be better at English after all.''}

A related issue concerned the representation of motion. In Group 9/7-B, a participant examining an intermediate image (Figure~\ref{fig:results-1-1-limitation}c) remarked that \textit{``if this airplane kept moving as it is, it might crash in three seconds.''} The comment illustrates how participants inferred possible movement from a static image and questioned the plausibility of the depicted scene.

Nevertheless, nonvisual ideas were not always lost in translation. In Group 9/7-A, participants imagined trees in a park producing music, and the generated image depicted musical notes as a visual proxy for sound (Figure~\ref{fig:results-1-1-limitation}d). This allowed the idea to remain recognizable despite the limitations of static imagery, as discussed further in Theme 3-2.

\paragraph{(ii) Selective materialization}
\begin{figure}[htb]
  \centering
  \includegraphics[keepaspectratio,width=1.0\linewidth]{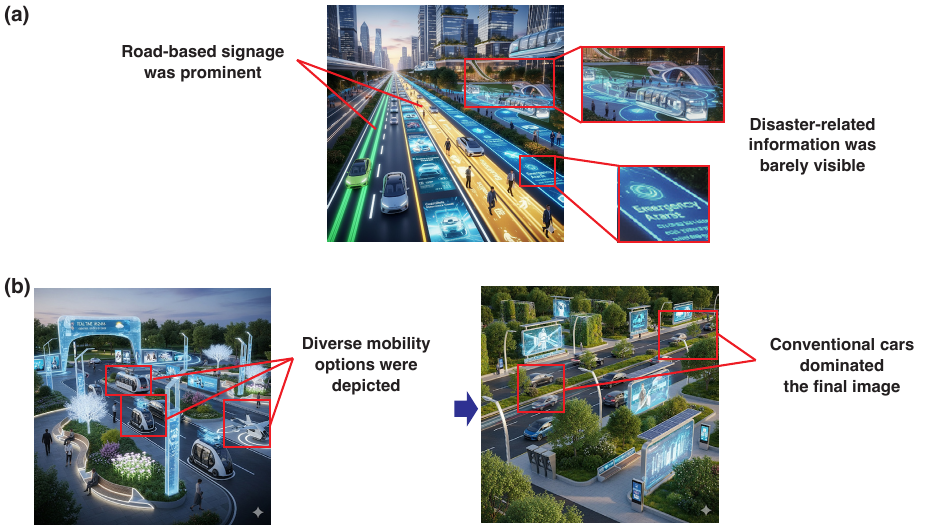}
  \caption{Theme 1-1(ii)--(iii): Examples of selective materialization and the loss of previously represented elements across image updates. (a) In Group 9/6-A, roads and mobility infrastructure were prominent, while disaster-related concerns included in the prompt were weakly represented in the generated image. (b) In Group 9/7-D, diverse forms of mobility depicted in an intermediate image were no longer evident in a subsequent image.}
  \label{fig:results-1-1-selection}
\end{figure}

Participants often proposed multiple concerns and possible applications of digital signage, which were subsequently combined into a single AI-generated image. In this process, some ideas became visually prominent, while others were weakly represented or omitted, even when they could potentially be visualized. This selectivity was particularly apparent when ideas concerning different functions, settings, or spatial scales were integrated into one image.

For example, in Group 9/6-A, several participants independently proposed applications of digital signage in disaster situations, alongside ideas regarding everyday information provision and future urban infrastructure:

\begin{itemize}
\item[PA-1:] \textit{Since parks can be gathering places during disasters, information about missing people or the arrival of essential supplies could be provided there.}
\item[PA-2:] \textit{I was thinking about roads in a disaster context. Normally multiple displays could present road signs or advertisements, but once an emergency occurs, they could switch to information about where to evacuate and how people should act.}
\item[PA-3:] \textit{I was thinking more about everyday information, such as providing notices about what is happening in a museum or a park that day.}
\item[PA-4:] \textit{Rather than conventional signage, the infrastructure itself could become signage---for example, roads could change color to guide people or vehicles toward particular routes.}
\item[PA-5:] \textit{For disasters, I also imagined a system in which people could actively obtain the specific information they need, rather than simply receiving the same information passively.}
\end{itemize}

Despite the prominence of disaster-related concerns in the initial discussion, these scenarios were barely visible in the generated image, which instead emphasized roads and mobility infrastructure (Figure~\ref{fig:results-1-1-selection}a). Notably, the disaster-related ideas had been explicitly included in the prompt through instructions such as \textit{``use information about missing persons''} and \textit{``provide information about the arrival of relief supplies.''}

\paragraph{(iii) Loss of previously represented elements across image updates}
Selective materialization also occurred across successive rounds of image updating. As facilitators revised prompts in response to group discussions, elements depicted in earlier images were sometimes reduced in prominence or absent from subsequent images. Thus, ideas that had once been visualized were not necessarily preserved throughout the iterative generation process.

A representative example emerged in Group 9/7-D, where participants initially emphasized the coexistence of diverse forms of mobility in future cities. One participant referred to emerging mobility services and envisioned roads accommodating \textit{``many different kinds of vehicles moving at different speeds.''} This diversity was visibly represented in an intermediate image (Figure~\ref{fig:results-1-1-selection}b, left). However, in a subsequent image (Figure~\ref{fig:results-1-1-selection}b, right), the variety of mobility options was no longer apparent, and the scene was dominated by conventional cars resembling those on contemporary streets. The group also did not mention this diversity when presenting its final ideas to the other participants.

This episode illustrates how successive image updates could result in the loss of previously represented elements, even when those elements had been central to participants' earlier ideas. Such changes in visual representation did not necessarily reflect an explicit decision to abandon these concerns, but could affect which aspects of the envisioned future remained available as visual references for subsequent discussion. Whether these concerns continued to be pursued in the discussion is examined in Theme 3.

\subsubsection{Theme 1-2: AI-generated supplementation}
\begin{figure}[htb]
\centering
\includegraphics[keepaspectratio,width=1.0\linewidth]{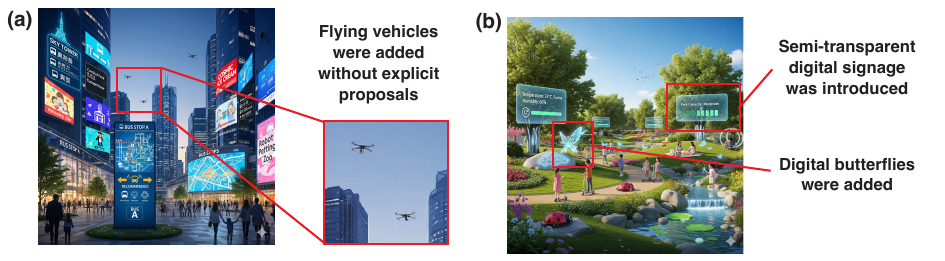}
\caption{Theme 1-2: Examples of AI-generated supplementation. (a) In Group 9/7-C, drone-like flying vehicles appeared despite not being explicitly proposed by participants. (b) In Group 9/6-C, the generated image included semi-transparent digital signage and digitally rendered butterflies.}
\label{fig:results-1-2-addition}
\end{figure}

AI-generated images did not simply reproduce participants' envisioned futures; they sometimes introduced visual elements that had not been explicitly proposed or represented existing ideas in unexpected forms. Examples included drone-like flying vehicles (Figure~\ref{fig:results-1-2-addition}a, Group 9/7-C), digitally rendered butterflies, and semi-transparent digital signage (Figure~\ref{fig:results-1-2-addition}b, Group 9/6-C). In Group 9/7-C, for instance, participants had not proposed flying vehicles explicitly, yet these appeared in the generated image as a concrete elaboration of their ideas.

However, the appearance of AI-generated additions did not necessarily lead to their incorporation into participants' ideas. Some additions became resources for further exploration, while others were merely noticed or received little attention. Thus, AI-generated supplementation introduced possible directions for discussion, but whether these were taken up depended on how participants engaged with the images, as examined in Themes 2 and 3.

\subsubsection{Theme 1-3: Organizational conditions of the image-centered workshop}
Beyond the content of individual images, the workshop format itself shaped the conditions under which participants envisioned future cities. Participants were asked to envision urban environments incorporating digital signage and to develop their visions through successive rounds of discussion and image updating. This premise was explicitly recognized in Group 9/6-C, when a participant asked, \textit{``So, are we supposed to imagine a future city in which signage is used?''} The question illustrates how participants oriented their ideas to the framing of the activity itself.

The timing of prompt submission and image generation introduced another constraint. Facilitators manually summarized group discussions into prompts and submitted them to the generative AI system, while participants often continued discussing their ideas as they waited for the next image. Contributions made after a prompt had been submitted could not be reflected in the image being generated unless they were reintroduced in a later round.

\subsection{Theme 2: Interpreting AI-generated images as provisional representations}
Building on Theme 1's account of selective materialization, this theme captures how participants and facilitators interpreted AI-generated images as provisional representations. Participants responded to the images in different ways, ranging from noticing unexpected elements and proposing revisions to questioning the depicted futures. Facilitators mediated these interactions by helping participants interpret the images and distinguish their own contributions from elements apparently introduced through image generation. We examine these different forms of engagement with AI-generated images (Theme 2-1) and the facilitators' roles in mediating image interpretation (Theme 2-2).

\begin{figure}[htb]
  \centering
  \includegraphics[keepaspectratio,width=1.0\linewidth]{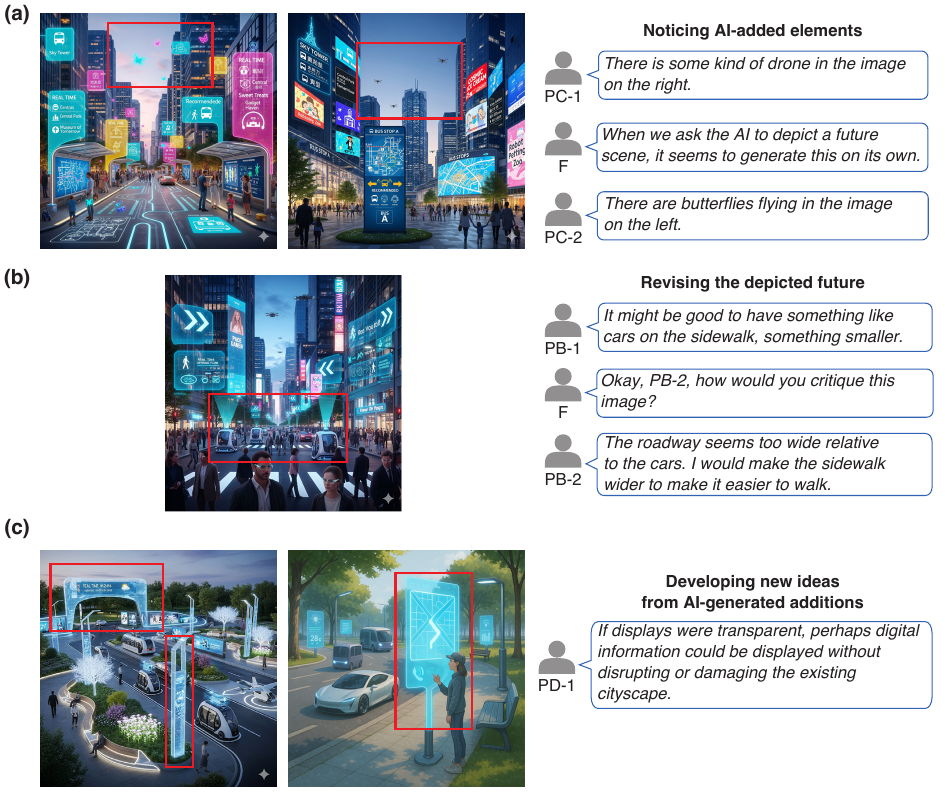}
  \caption{Theme 2-1(i): Examples of revising and developing ideas from AI-generated images. (a) In Group 9/7-C, participants noticed unexpected elements, including drones and butterflies, while the facilitator suggested that these had been introduced by the AI. (b) In Group 9/6-B, participants proposed revisions to the depicted roadway, sidewalks, and vehicles. (c) In Group 9/7-D, semi-transparent digital signage prompted a new idea for displaying information without disrupting the existing cityscape. }
  \label{fig:results-2-1-negotiation}
\end{figure}

\subsubsection{Theme 2-1: Revising, developing, and resisting provisional images}
\paragraph{(i) Revising and developing ideas from AI-generated images}
In Group 9/7-C, participants examined details in the generated images and recognized elements that had not been explicitly proposed in the preceding discussion (Figure~\ref{fig:results-2-1-negotiation}a):

\begin{itemize}
\item[PC-1:] \textit{There is some kind of drone in the image on the right.}
\item[F:] \textit{Yes, there is a drone. When we ask the AI to depict a future scene, it seems to generate this on its own, although I am not really sure why.}
\item[PC-2:] \textit{There are butterflies flying in the image on the left.}
\item[F:] \textit{Yes. I am not sure how that would actually be realized physically, though.}
\item[PC-1:] \textit{Interesting. I have never seen drone-based signage before.}
\end{itemize}

This exchange illustrates how participants recognized and considered unexpected visual elements, with the facilitator helping to distinguish AI-generated additions from their own proposals. The participants' responses also suggest that these additions could attract curiosity.

In Group 9/6-B, whose members were primarily elementary-school-aged children, participants similarly examined specific features of the generated image but focused on how the depicted environment could be improved (Figure~\ref{fig:results-2-1-negotiation}b):

\begin{itemize}
\item[PB-1:] \textit{There are things like cars on the roadway, but it might be good to have something like that on the sidewalk too, something smaller.}
\item[F:] \textit{So, you mean smaller cars on the sidewalk?}
\item[PB-1:] \textit{Yes, something smaller.}
\item[F:] \textit{Okay. PB-2, how would you critique this image?}
\item[PB-2:] \textit{This is more of a physical issue, but the roadway seems too wide relative to the cars. I would make the sidewalk wider to make it easier to walk.}
\end{itemize}

Here, the generated image provided a visual reference for articulating possible revisions. By examining the relative sizes of vehicles, roads, and sidewalks, participants identified spatial relationships they wished to change and proposed alternative arrangements.

In other cases, participants developed unexpected visual elements into new design ideas. In Group 9/7-D, for example, an AI-generated image depicted semi-transparent digital signage, although participants had previously proposed compact holographic displays rather than explicitly requesting this visual feature (Figure~\ref{fig:results-2-1-negotiation}c). One participant drew on this addition to articulate a new design rationale: \textit{``One thing I noticed in both images is that the displays are semi-transparent, almost transparent. If they were transparent, perhaps digital information could be displayed without disrupting or damaging the existing cityscape.''} In this case, a feature introduced through image generation became a resource for considering how digital information could be integrated into urban environments while preserving their visual landscape.

\paragraph{(ii) Multiple forms of resistance to AI-generated images}
\begin{figure}[htb]
  \centering
  \includegraphics[keepaspectratio,width=1.0\linewidth]{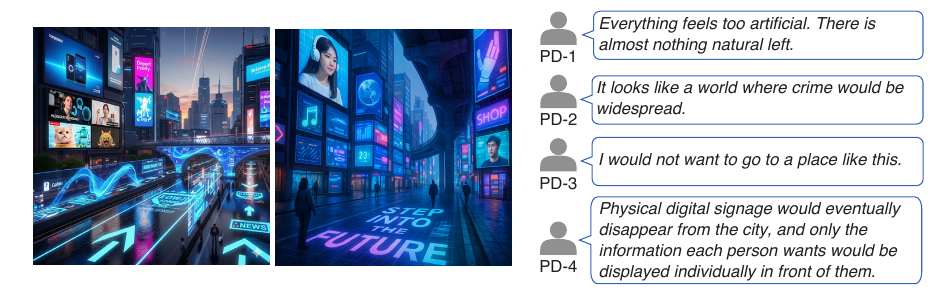}
  \caption{Theme 2-1(ii): Examples of questioning the desirability of an AI-generated future cityscape. In Group 9/6-D, the images depict urban environments extensively incorporating digital signage, based on ideas proposed during the discussion. The accompanying excerpts illustrate participants' discomfort with the visualized cityscape and their consideration of alternative ways to provide digital information.}
  \label{fig:results-2-1-resistance}
\end{figure}
Participants did not always accept AI-generated images as adequate representations of their ideas or as desirable visions of the future. We observed two contrasting forms of resistance: one arose when images failed to adequately represent participants' intentions, while the other emerged when participants saw their own proposals materialized in an image and questioned the desirability of the resulting future. The former was evident in Group 9/6-C, as described in Theme 1-1(i), where participants questioned the limited representation of auditory guidance and text-based information (Figure~\ref{fig:results-1-1-limitation}a--b).

A different form of resistance emerged in Group 9/6-D. During initial ideation, several participants proposed extending digital displays beyond conventional screens to urban infrastructure, including bridges, road surfaces, and guardrails. The resulting AI-generated images combined these ideas into urban scenes (Figure~\ref{fig:results-2-1-resistance}). However, seeing such environments, participants questioned whether they would actually want to live in them:

\begin{itemize}
\item[PD-1:] \textit{In this image, everything feels too artificial. There is almost nothing natural left. [...] If everything were flashing whenever you looked up, I think it would become visually exhausting.}
\item[PD-2:] \textit{At first glance, it looks like a world where crime would be widespread. I would want something brighter that does not come from digital light—more greenery or sunlight.}
\item[PD-3:] \textit{I would not want to go to a place like this, although I can imagine that some people designing such systems might want to display information everywhere.}
\item[PD-4:] \textit{My thinking has changed through this discussion. For people who want the latest technology and people who prefer a more natural environment to coexist, perhaps physical digital signage would eventually disappear from the city, and only the information each person wants would be displayed individually in front of them.}
\end{itemize}

Unlike Group 9/6-C, where participants questioned whether the generated images adequately represented their intentions, participants in Group 9/6-D questioned the future that emerged when their own proposals were brought together in a single cityscape. Seeing these proposals materialized encouraged participants to reconsider whether such an extensively digitized environment would be desirable. PD-4 explicitly described a change in perspective and proposed an alternative in which information would be presented individually. Thus, visualizing participants' own proposals could prompt them to reconsider the desirability of the future those proposals collectively implied.

\subsubsection{Theme 2-2: Facilitating image interpretation and clarifying contributions}
Across the groups, facilitators supported participants in interpreting AI-generated images by identifying how their ideas had been reflected, pointing out discrepancies, and requesting further image generation when necessary. They also helped participants distinguish elements of earlier discussions from those introduced through the generative process. For example, in Group 9/7-C, the facilitator suggested that an unexpected drone had been generated by the AI \textit{``on its own''} (Figure~\ref{fig:results-2-1-negotiation}a).

However, identifying discrepancies did not necessarily lead to the reconsideration of concerns that were weakly represented. In Group 9/6-A, a participant questioned where the park discussed earlier was depicted in the generated image (Figure~\ref{fig:results-1-1-selection}a). The facilitator pointed to a possible representation but expressed uncertainty about whether it adequately reflected the group's proposal. Subsequent discussion developed primarily around roads and mobility infrastructure, while the earlier park-related concerns received limited attention.

Beyond facilitators' efforts, participants themselves also reflected on the relationship between their contributions and the generative process. In Group 9/7-D, for example, participants reflected on their experience of using generative AI during the workshop:

\begin{itemize}
\item[F:] \textit{How was it to use generative AI in this workshop? I imagine many of you have not often used it as part of a discussion like this.}
\item[PD-1:] \textit{I realized that it can easily visualize some things, but it also fills in missing information on its own. I think there are both advantages and disadvantages to that.}
\item[PD-2:] \textit{I also wonder how much the result is constrained by the kind of future image the AI produces even when only basic information is provided—how much reflects human tendencies and how much comes from tendencies in the AI itself.}
\end{itemize}

These reflections suggest that participants themselves recognized the selective and imperfect nature of AI-generated representations (Theme 1). They acknowledged that AI could fill in missing details and questioned how its own tendencies might shape the resulting images.

\FloatBarrier

\subsection{Theme 3: Evolving concerns and uneven attention in future visions}
Building on Themes 1 and 2, this theme captures how participants' concerns evolved, received less attention, or persisted in new forms across successive rounds of discussion and AI-image generation. Across groups, discussions often elaborated the visual and spatial features of future cities, while the purposes, situations of use, and infrastructural implications of proposed technologies received less sustained attention. We identified three dynamics: (i) elaboration of visual and spatial aspects (Theme 3-1), (ii) less sustained exploration of purposes, information content, and infrastructural conditions (Theme 3-2), and (iii) persistence and rearticulation of underlying concerns (Theme 3-3). Detailed trajectories for all eight groups are provided in Appendix~\ref{app:workshop-trajectories} (Table~\ref{tab:theme3-cross-case} and Figures~\ref{fig:results-3-chart0906A}--\ref{fig:results-3-chart0907D}).

\subsubsection{Theme 3-1: Elaborating visual and spatial aspects of future cities}
\begin{figure}[htb]
  \centering
  \includegraphics[keepaspectratio,width=1.0\linewidth]{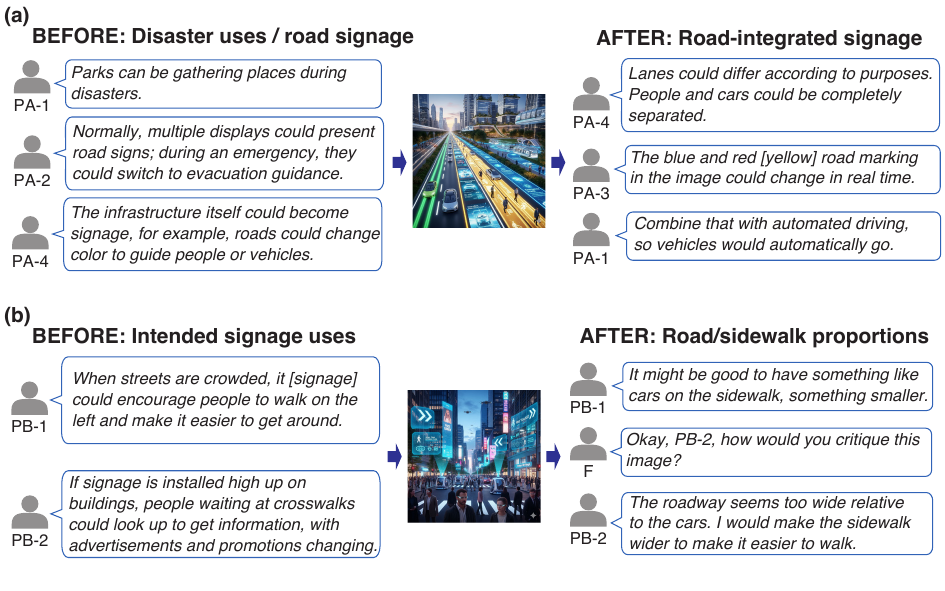}
  \caption{Theme 3-1: Examples of the elaboration of visual and spatial aspects of future cities in (a) Group 9/6-A and (b) Group 9/6-B. The excerpts do not necessarily represent consecutive turns. ``Before'' and ``After'' refer to discussions before and after the first group-specific image update.}
  \label{fig:results-3-1-before-after}
\end{figure}

Image-based discussions often developed existing proposals into more concrete spatial and visual designs. Two prominent directions concerned (i) mobility and sites of implementation and (ii) urban environments and landscapes.

\paragraph{(i) Mobility and sites of implementation.}
Initial ideas about mobility and signage locations were often elaborated through image-based discussion. In Group 9/6-A, initial proposals ranged from disaster-related information provision in parks and evacuation guidance to roads that could change color to direct traffic (Figures~\ref{fig:results-3-1-before-after}a and \ref{fig:results-3-chart0906A}). The generated image prominently depicted road-integrated signage, whereas disaster-related applications were barely visible despite their inclusion in the prompt (Theme 1-1(ii)). Subsequent discussions developed the road-based proposal through lane arrangements, pedestrian--vehicle separation, and real-time road markings (Figures~\ref{fig:results-3-1-before-after}a and \ref{fig:results-3-chart0906A}).

Similarly, in Group 9/7-B, proposals for displaying information on aircraft and other structures expanded to include robots and overhead displays (Figure~\ref{fig:results-3-chart0907B}). In Group 9/7-A, an idea for tree-integrated signage developed into discussions of display shapes and installation methods (Figure~\ref{fig:results-3-chart0907A}). These cases illustrate how initial proposals concerning signage locations and display surfaces were expanded or refined into more concrete configurations.

\paragraph{(ii) Visual environments and urban landscapes}
Discussions also shifted toward the appearance and spatial arrangements of future urban environments. In Group 9/6-B, participants initially proposed applications such as pedestrian guidance and advertising. After viewing the updated image, however, they examined the depicted street layout, suggesting smaller vehicles for sidewalks and questioning the relative widths of roads and sidewalks (Theme 2-1(i); Figures~\ref{fig:results-3-1-before-after}b and \ref{fig:results-3-chart0906B}). Here, the discussion shifted from the functions of digital signage to the spatial arrangement of the street, while the underlying concern with pedestrian movement continued through proposals for improving walkability.

Attention to the depicted urban environment could also extend beyond spatial refinement to the desirability of the envisioned future. In Group 9/6-D, an image depicting an extensively digitized city led participants to question whether such an environment would be desirable (Theme 2-1(ii); Figure~\ref{fig:results-3-chart0906D}). Subsequent discussions explored urban comfort and the coexistence of digital technologies and nature. Thus, discussions of visual environments involved reconsidering the kinds of urban landscapes participants wished to create.

\subsubsection{Theme 3-2: Less sustained exploration of purposes, information content, and infrastructural conditions}
\begin{figure}[htb]
  \centering
  \includegraphics[keepaspectratio,width=1.0\linewidth]{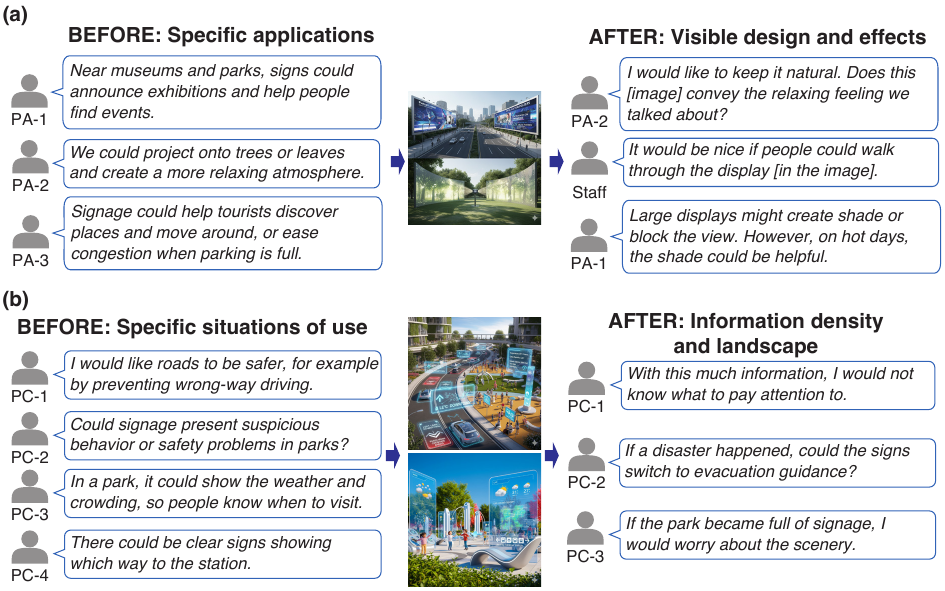}
  \caption{Theme 3-2: Examples of less sustained exploration of specific purposes and situations of use in (a) Group 9/7-A and (b) Group 9/6-C. The excerpts do not necessarily represent consecutive turns. ``Before'' and ``After'' refer to discussions before and after the first group-specific image update.}
  \label{fig:results-3-2-before-after}
\end{figure}

While participants elaborated visual and spatial designs, other concerns were not always explored to the same extent. This tendency was evident in three areas: (i) purposes and situations of use, (ii) information content and communication modes, and (iii) infrastructural conditions and consequences.

\paragraph{(i) Purposes and situations of use}
Participants initially proposed various applications of digital signage for specific situations, but subsequent discussions often shifted toward refining the depicted environments. In Group 9/7-A, initial proposals included event information, tourism support, congestion relief, and relaxing public spaces (Figures~\ref{fig:results-3-2-before-after}a and \ref{fig:results-3-chart0907A}). Following the image update, participants focused on whether the depicted environment conveyed a relaxing atmosphere and how the proposed displays might affect movement, views, and shade. Although the goal of creating a relaxing public space persisted, specific applications such as tourist guidance and congestion relief received less attention. A similar shift occurred in Group 9/6-B, where initial proposals for advertising and pedestrian guidance gave way to discussions of street proportions, display sizes, and lighting (Theme 3-1(ii); Figure~\ref{fig:results-3-chart0906B}). These cases illustrate how the purposes motivating initial proposals could recede as participants developed their spatial and visual configurations.

\paragraph{(ii) Information content and communication modes}
A related tendency concerned what information digital signage would convey and how it would be communicated. In Group 9/6-C, participants initially proposed safety warnings and everyday information provision (Figures~\ref{fig:results-3-2-before-after}b and \ref{fig:results-3-chart0906C}). After viewing the generated images, they questioned whether the images adequately represented sound, Japanese text, and safety-related information (Themes 1-1(i) and 2-1(ii)). Subsequent discussions addressed information density, evacuation guidance, and the visual impact of signage on park scenery, while auditory guidance and Japanese-language information received little further exploration.

Similar patterns appeared elsewhere. Multilingual guidance in Group 9/7-B and voice-responsive guidance in Group 9/7-D received limited follow-up as discussions developed around visible design features (Figures~\ref{fig:results-3-chart0907B} and \ref{fig:results-3-chart0907D}). Together, these cases illustrate how specific information content and alternative communication modes received less sustained attention than the visual and spatial design of digital signage.

Nevertheless, nonvisual ideas were not always lost. In Group 9/7-A, a proposal for trees producing music was represented symbolically by musical notes in the generated images (Theme 1-1(i); Figure~\ref{fig:results-3-chart0907A}). Although this representation could not convey sound itself, it preserved the idea as a recognizable element of the group's evolving discussion.

\paragraph{(iii) Infrastructural conditions and consequences}
Participants also raised practical requirements and broader influences of future technologies, although these concerns were generally discussed only intermittently. In Group 9/7-A, participants considered the effects of shade from large displays and the possibility of solar power generation (Figure~\ref{fig:results-3-chart0907A}). In Group 9/7-D, discussions addressed emergency power supply and potential collisions with semi-transparent displays (Figure~\ref{fig:results-3-chart0907D}). Group 9/6-C also raised concerns about excessive surveillance and mistaken identification (Figure~\ref{fig:results-3-chart0906C}). Although some concerns, such as emergency power supply in Group 9/7-D, were revisited and included in the final presentation, they rarely developed into sustained lines of discussion. Across these cases, visible design features were elaborated more consistently than operational requirements and broader social implications.

\subsubsection{Theme 3-3: Persistence and rearticulation of underlying concerns}
Alongside these shifts in discussion focus, some groups continued to develop earlier concerns through different proposals. In Group 9/6-A, although specific disaster-related applications receded, an underlying concern with context-dependent information provision persisted through proposals for real-time traffic guidance (Figure~\ref{fig:results-3-chart0906A}). Similarly, in Group 9/6-C, initial concerns about safety and everyday information provision developed into discussions of personalization, appropriate information density, and responsiveness to different situations (Figure~\ref{fig:results-3-chart0906C}). In Group 9/7-C, concerns about excessive visual information were rearticulated through proposals for selectively accessible AR-based guidance (Figure~\ref{fig:results-3-chart0907C}).

These cases show that participants could sustain and reconsider underlying concerns through different proposals, even when the immediate topics of discussion changed. Thus, receiving less attention in a particular form did not necessarily indicate that a concern had been completely abandoned.

\subsection{Theme 4: Reconfiguring participation and mediation around AI-generated images}
Beyond changes in discussion topics (Theme 3), AI-generated images also became involved in how participants interacted and developed ideas together. We examine three aspects: (i) reorganizing participation around AI-generated images (Theme 4-1), (ii) facilitation of discussion and image revision (Theme 4-2), and (iii) the involvement of other resources (Theme 4-3).

\subsubsection{Theme 4-1: Reorganizing participation around AI-generated images}
AI-generated images could reshape whose contributions became central to subsequent discussions by making some proposals more prominent. In Group 9/6-A, the generated image foregrounded one participant's proposal for road-integrated signage (Theme 1-1(ii); Figures~\ref{fig:results-3-1-before-after}a and \ref{fig:results-3-chart0906A}). The participant who had proposed road-integrated signage subsequently became central to discussions of traffic organization. By contrast, another participant questioned where the previously discussed park was depicted, asking \textit{``Is the park somewhere else?''}, but made no further recorded contributions. This contrast suggests that selective materialization could create unequal opportunities for participants' contributions to be taken up and developed.

However, less prominent initial proposals did not always limit their contributors' subsequent participation. In Group 9/6-A, a participant who had initially proposed disaster-related information joined the road-focused discussion by suggesting automated driving (Theme 3-3). Similarly, in Group 9/7-D, a participant whose initial proposal for voice-responsive guidance received little further attention later drew on the depicted displays' transparency to suggest preserving the cityscape (Theme 2-1(i); Figure~\ref{fig:results-3-chart0907D}). These cases illustrate how participants could continue contributing through emerging ideas, even when their initial proposals received less attention.

\subsubsection{Theme 4-2: Facilitators as mediators of discussion and image revision}
Facilitators organized participants' contributions by eliciting and clarifying ideas, connecting proposals, and summarizing discussions for image generation. In Group 9/6-B, the facilitator initially paraphrased the children's ideas. After presenting the generated image, the facilitator invited critiques, prompting participants to suggest changes to road and sidewalk proportions  (Theme 2-1(i); Figure~\ref{fig:results-3-chart0906B}). In Group 9/6-D, follow-up questions encouraged participants who rejected the depicted cityscape to elaborate alternative visions of desirable urban environments (Theme 2-1(ii); Figure~\ref{fig:results-3-chart0906D}).

Facilitators also managed the pace of interaction. In Group 9/7-B, where children rapidly proposed numerous ideas, the facilitator occasionally personified the AI to organize their contributions. When additional ideas continued to emerge, the facilitator remarked, \textit{``the AI is having quite a hard time,''} and asked participants to wait while the AI worked on the first image. Later, responding to the children's playful concern that the AI might be overwhelmed, the facilitator added, \textit{``We should be kind to the AI.''} By attributing difficulty and effort to the personified AI, the facilitator humorously acknowledged the children's contributions while managing the pace of discussion and image revision (Figure~\ref{fig:results-3-chart0907B}).

\subsubsection{Theme 4-3: Involvement of personal experiences, expertise, and other resources}
Participants also drew on personal experiences and knowledge to develop future visions. In Group 9/6-B, an earlier workshop demonstration inspired a proposal to guide pedestrian movement through psychological responses. In Group 9/7-C, knowledge of existing transportation technologies informed mobility-related proposals, while in Group 9/6-D, personal experiences and preferences provided grounds for questioning the desirability of the depicted cityscape (Theme 2-1(ii)).

Other workshop materials and technical expertise also supported discussion. In Group 9/7-D, revisiting an earlier worksheet prompted consideration of emergency power supply, while a workshop staff member with expertise in digital signage explained that unconventional display shapes could already be realized by combining projectors with flexible screens, connecting participants' ideas to existing technical possibilities (Figure~\ref{fig:results-3-chart0907D}). Similarly, in Group 9/6-C, the facilitator's technical explanations helped participants understand the limitations of representing sound and Japanese text (Theme 1-1(i); Figure~\ref{fig:results-3-chart0906C}). These cases show how experiences, expertise, and earlier materials contributed to the development of future visions alongside AI-generated images.

\section{Discussion}
Our analysis identified four interrelated dynamics in AI-supported collaborative design. The images selectively materialized participants' concerns while introducing unexpected elements (Theme 1), which participants interpreted, revised, or resisted as provisional representations (Theme 2). Through repeated cycles of discussion and image generation, some concerns were elaborated while others received less sustained attention (Theme 3). These processes also shaped participation and mediation through interactions among participants, facilitators, and other resources (Theme 4). We discuss these findings in relation to intermediary objects, the socio-technical dimensions of imagined futures, and the organization of participation in AI-supported cooperative design.

\subsection{AI-generated images as intermediary objects: Generative roles and unintended selectivity}
\label{sec:discussion-theme1-2}
Prior research has shown that intermediary objects, including sketches, drawings, and digital models, do more than reproduce designers' ideas: they mediate how ideas are externalized and developed through collaborative work~\cite{vinck1994mediating,vinck2012accessing,paavola2019dynamics}. Their generative roles have been described in terms of ``branching out,'' which opens new possibilities, and ``unfolding,'' which supports the elaboration of problems and solutions \cite{sadorge2024generative}. AI-generated images in our workshop performed similar functions. Unexpected visual elements sometimes prompted new ideas, while visualizing participants' own proposals encouraged them to reconsider the desirability of the futures they envisioned (Themes 1 and 2).

However, one difference concerned the additional interpretation of generated images. Because the transformation from discussions and prompts into images was not fully controllable, participants and facilitators had to assess how earlier ideas had been represented. While previous studies identified difficulties in translating design intentions into generative images~\cite{verheijden2023collaborative} and the conversational value of imperfect images~\cite{guridi2024fake,guridi2025image}, our findings reveal more varied responses: participants accepted and elaborated unexpected additions, questioned inadequate representations, or reconsidered the desirability of futures that materialized their own proposals (Theme 2). Rather than serving as transparent records of participants' ideas, AI-generated images themselves became objects of interpretation and negotiation.

Another important difference concerns how AI-generated images selectively materialize participants' concerns. Although conventional intermediary objects also involve selective abstraction~\cite{vinck2012accessing}, AI-generated images can introduce selections that participants and facilitators did not deliberately make. In Group 9/6-A, for example, disaster-related applications raised by multiple participants were included in the prompt, yet the generated image foregrounded one participant's proposal for road-based signage (Theme 1-1(ii)). Previously depicted elements could also disappear across image revisions without explicit decisions to exclude them (Theme 1-1(iii)). Such transformations could shift subsequent discussions, with initially salient concerns receiving less attention without being explicitly rejected (Section~\ref{sec:discussion-theme3}).

\subsection{Visual concretization and the backgrounding of infrastructural conditions}
\label{sec:discussion-theme3}
Previous research has shown that generative AI can support design exploration while also encouraging reliance on generated features and design fixation~\cite{cai2023designaid,paananen2024using,wadinambiarachchi2024effects,ge2025effects}. These effects are not necessarily contradictory~\cite{crilly2017next}. Our process-oriented analysis suggests that such expansion and narrowing can coexist within collaborative discussions. AI-generated images elaborated the visual and spatial configurations of future cities (Theme 3-1), while concerns about specific purposes, situations of use, and nonvisual forms of information were less consistently sustained (Theme 3-2). Thus, AI-generated images could redistribute attention among different aspects of the envisioned futures, rather than simply expanding or constraining discussion topics.

Such shifts are particularly consequential when envisioning technologies embedded in public environments~\cite{kubota2026nudging,guridi2024fake,awashra2025generative}, where their significance depends on existing infrastructures, everyday practices, and socio-technical arrangements~\cite{star1996steps,neumann1996making}. Imagining futures as if they emerged from a blank slate risks overlooking these conditions and the constraints shaping possible technological change~\cite{selin2015against,selin2017experiments}. An infrastructural perspective therefore calls for considering not only how future technologies might appear, but also whom they would serve, how they would operate and be maintained, and what social and environmental consequences they might entail~\cite{wong2020infrastructural}.

Viewed through this infrastructural perspective, our findings point to a tension in using AI-generated images to envision urban futures. The images enabled participants to explore new possibilities through shared visual references. On the other hand, successive revisions tended to elaborate what these futures might look like more extensively than how the envisioned technologies could be used, operated, or maintained. While such visual exploration is valuable for open-ended speculation, this imbalance presents a limitation for civic engagement and cooperative design that seek to consider how future technologies could plausibly be integrated into existing urban environments~\cite{selin2015against,wong2020infrastructural}.

However, these developments were not determined by AI-generated images alone. Participants' experiences, technical expertise, and earlier workshop materials sometimes brought less visible socio-technical concerns back into discussion (Theme 4-3). Although these resources did not often sustain such concerns, they suggest that other forms of knowledge and mediation could counterbalance the emphasis on visualized features.

\subsection{Reconfiguring participation through AI-generated images}
The implications of AI-generated images extend beyond what is discussed to how participation itself is organized. Relational perspectives on participation emphasize that participants' roles and opportunities for engagement are not fixed in advance but emerge through interactions among people, artifacts, and other elements of collaborative practice~\cite{andersen2015participation,chilvers2020remaking,pedersen2020staging,pedersen2022navigating}. From this perspective, AI-generated images could become part of these interactions, reshaping how participants and facilitators engaged with one another and contributed to the collaborative design process.

One implication was a change in how participants engaged in the design process. Participants could also contribute by interpreting unexpected elements, critiquing the depicted environments, or questioning the desirability of the resulting futures (Themes 2 and 4-2). Facilitators helped turn these responses into further discussion and image revisions. Previous research has highlighted their role in supporting conversations around imperfect AI-generated images~\cite{guridi2025image}. Our findings further suggest that participation involved not only proposing what should be visualized but also critically examining and revising the images that emerged.

A further implication concerns how AI-generated images may unevenly shape opportunities for participation. In our workshop, some proposals became visually prominent without an explicit collective decision to prioritize them (Theme 4-1). This prominence could reinforce the proposers' roles in subsequent discussions, while giving others fewer opportunities to develop their own ideas. Although participants could still contribute through emerging topics, such shifts raise concerns about unequal participation. Previous research has emphasized that opportunities to speak do not necessarily ensure equal influence over collective decisions~\cite{davies2013rules,vines2013configuring,bratteteig2016unpacking}. Our findings suggest that AI-generated images may contribute to these inequalities by shaping whose contributions become central to collective exchanges.

\subsection{Toward accessible and reflective AI-mediated cooperative design}
Generative AI can make visual exploration accessible without specialized visualization skills. In our workshop, participants, including elementary-school-aged children, actively discussed and revised generated images. Questionnaire responses also indicated that the images supported idea sharing and collaboration (Appendix~\ref{app:questionnaire-results}). However, as our analysis suggests, image generation could also selectively foreground concerns and reshape participation. The challenge is therefore to preserve the benefits of image generation while supporting reflection on how it shapes collaborative work.

One approach is to introduce brief, structured reflection between image revisions. Although our worksheet before the group exchanges encouraged participants to consider different uses and values, these concerns were not always sustained in subsequent discussions. Previous studies have used discussion cards to encourage participants to consider various perspectives and topics~\cite{felt2018imagine,pedersen2025staging}. Similar activities could prompt participants to consider who would use and maintain the proposed technologies, which concerns are missing from the current image, and which ideas should be carried forward. Such reflection could help groups revisit overlooked concerns and negotiate priorities before generating the next image.

Another approach is to alternate speculative image generation with activities grounded in existing places and practices. In a previous study, participants photographed their city and used these photographs to reflect on its past, present, and possible futures~\cite{selin2017experiments}. Similar observation activities could connect AI-generated visions to everyday experiences and existing infrastructures. Combining different forms of representation could also help participants explore socio-technical relationships that are difficult to capture in a single image.

Future AI tools could be developed to support these reflective practices without placing additional demands on facilitators. Such tools could analyze discussion records, prompts, and successive images to identify concerns that were omitted, transformed, or received limited attention. They could then suggest context-specific questions or reflection cards to help participants revisit these concerns before further image revisions. By combining accessible image generation with support for reflection on evolving discussions, such tools could make AI-mediated cooperative design both more accessible and more reflexive.

\subsection{Limitations and future work}
This study has several limitations. First, it examined a single workshop format focused on future urban environments and digital signage. The preceding demonstrations and preparatory activities may also have shaped participants' ideas and subsequent discussions. For example, a participant drew on the demonstration experience when proposing pedestrian guidance. Participants also showed relatively high interest in science and technology. Our findings may therefore differ across workshop contexts, participant groups, and facilitation practices.

Second, the workshop was not designed as a hypothesis-testing experiment and did not include a control condition. We therefore cannot fully isolate the effects of AI-generated images from those of facilitation, other workshop materials, and group dynamics. Our process-oriented analysis instead identifies patterns that can inform future research directions. Comparative and controlled studies could examine how different forms of visualization and workshop procedures shape the development of concerns and opportunities for participation.

Third, some observed difficulties reflected limitations of the image-generation systems used in this study, particularly in representing sound, motion, and written text. Advances in multimodal generation may reduce these constraints. However, greater generative capabilities would not necessarily eliminate the selective transformation of participants' concerns into shared representations. Such selectivity could continue to influence which ideas receive attention and how participation develops. Future research should therefore examine how generative AI shapes collaborative processes, alongside efforts to improve its technical capabilities. These insights could guide the development of novel technologies that mitigate unintended effects of AI support on discussion and participation while enhancing the benefits of AI-supported creative collaboration.

\section{Conclusions}
This study investigated how multiparty cooperative design evolved through repeated cycles of AI-image generation and discussion. Our process-oriented qualitative analysis identified four interrelated themes: (1) AI-generated images selectively materialized participants' concerns and introduced unexpected elements; (2) participants interpreted, revised, or resisted these images as provisional representations; (3) discussions elaborated visual and spatial aspects of future cities, while purposes, situations of use, and infrastructural conditions received less sustained attention, sometimes without explicit rejection or decisions to deprioritize them; and (4) participation evolved through interactions among participants, facilitators, images, and other resources.

Our findings extend the understanding of AI-generated images as intermediary objects in cooperative design. Like conventional design artifacts, these images can support the development and critical reconsideration of ideas while selectively representing participants' concerns. However, the selection introduced through AI generation may diverge from participants' expressed priorities. Such unanticipated selections can foreground particular proposals without collective agreement, influencing both the direction of discussion and whose contributions gain prominence.

These findings do not negate the benefits of AI-generated images, which made visual exploration accessible without specialized skills in our workshop. Rather, they suggest the importance of providing opportunities to reflect on how collective visions evolve through image revision. Structured activities, such as reflection cards and engagement with existing places, could help participants reconsider overlooked concerns about intended users, operational requirements, and broader consequences. AI-based tools could further support this process by tracing concerns across discussions, prompts, and successive images and suggesting questions for collective reflection.

\section*{Acknowledgment}
We would like to express our sincere gratitude to the staff of the National Museum of Emerging Science and Innovation (Miraikan) for their invaluable support in organizing and conducting the workshop. This work was supported by Japan Science and Technology Agency [JPMJPR25IA].
\bibliographystyle{ACM-Reference-Format}
\bibliography{ref}

@String{Computing = "Computing" }

@String{Computer = "{IEEE} Computer" }

@String{Springer = "Springer-Verlag" }

@article{heigl2026generative,
  title={Generative artificial intelligence in creative contexts: a systematic review and future research agenda},
  author={Heigl, Rebecca},
  journal={Management Review Quarterly},
  volume={76},
  number={1},
  pages={955--992},
  year={2026},
  publisher={Springer}
}

@inproceedings{chiou2023designing,
  title={Designing with AI: An exploration of co-ideation with image generators},
  author={Chiou, Li-Yuan and Hung, Peng-Kai and Liang, Rung-Huei and Wang, Chun-Teng},
  booktitle={Proceedings of the 2023 ACM designing interactive systems conference},
  pages={1941--1954},
  year={2023}
}

@article{rafner2023creativity,
  title={Creativity in the age of generative AI},
  author={Rafner, Janet and Beaty, Roger E and Kaufman, James C and Lubart, Todd and Sherson, Jacob},
  journal={Nature Human Behaviour},
  volume={7},
  number={11},
  pages={1836--1838},
  year={2023},
  publisher={Nature Publishing Group UK London}
}

@article{wu2026human,
  author    = {Wu, Xingting and Tran, Tram Thi Minh
               and Zhou, Qiqi and Sun, Jian and Wu, Hao
               and Wang, Xiaoyang},
  title     = {Human-{AI} co-creation or conflict?
               Mapping art students' diverse perspectives
               on creative identity with {genAI}:
               A Q methodology study},
  journal   = {Education and Information Technologies},
  volume    = {31},
  number    = {9},
  pages     = {2837--2876},
  year      = {2026},
  doi       = {10.1007/s10639-026-13914-4}
}

@article{fu2020research,
  title={Research on human--AI co-creation based on reflective design practice: Z. Fu, Y. Zhou},
  author={Fu, Zhiyong and Zhou, Yuyao},
  journal={CCF Transactions on Pervasive Computing and Interaction},
  volume={2},
  number={1},
  pages={33--41},
  year={2020},
  publisher={Springer}
}

@article{lee2024empirical,
  title={An empirical investigation of the impact of ChatGPT on creativity},
  author={Lee, Byung Cheol and Chung, Jaeyeon},
  journal={Nature Human Behaviour},
  volume={8},
  number={10},
  pages={1906--1914},
  year={2024},
  publisher={Nature Publishing Group UK London}
}

@article{ge2025effects,
  title={The effects of generative AI model type and visual stimuli type on design creativity},
  author={Ge, Weichu and Hou, Guanhua},
  journal={AI EDAM},
  volume={39},
  pages={e17},
  year={2025},
  publisher={Cambridge University Press}
}

@article{wei2025effects,
  title={The effects of generative AI on collaborative problem-solving and team creativity performance in digital story creation: an experimental study},
  author={Wei, Xiaodong and Wang, Lei and Lee, Lap-Kei and Liu, Ruixue},
  journal={International Journal of Educational Technology in Higher Education},
  volume={22},
  number={1},
  pages={23},
  year={2025},
  publisher={Springer}
}

@article{michel2025collaborative,
  title={Collaborative writing based on generative AI models: Revision and deliberation processes in German as a foreign language},
  author={Michel, Marije and Bazhutkina, Iryna and Abel, Niklas and Strobl, Carola},
  journal={Journal of Second Language Writing},
  volume={67},
  pages={101185},
  year={2025},
  publisher={Elsevier}
}

@inproceedings{he2024revealing,
  author    = {He, Wenhui and Xiao, Yi and Xie, Yu},
  title     = {Revealing user tacit knowledge:
               Generative-Image-{AI} helps create
               better design conversation},
  booktitle = {DRS2024: Boston},
  year      = {2024},
  doi       = {10.21606/drs.2024.1329}
}

@article{youmans2014design,
  title={Design fixation: Classifications and modern methods of prevention},
  author={Youmans, Robert J and Arciszewski, Thomaz},
  journal={AI EDAM},
  volume={28},
  number={2},
  pages={129--137},
  year={2014},
  publisher={Cambridge University Press}
}

@article{crilly2017next,
  title={Where next for research on fixation, inspiration and creativity in design?},
  author={Crilly, Nathan and Cardoso, Carlos},
  journal={Design Studies},
  volume={50},
  pages={1--38},
  year={2017},
  publisher={Elsevier}
}

@inproceedings{cai2023designaid,
  title={DesignAID: Using generative AI and semantic diversity for design inspiration},
  author={Cai, Alice and Rick, Steven R and Heyman, Jennifer L and Zhang, Yanxia and Filipowicz, Alexandre and Hong, Matthew and Klenk, Matt and Malone, Thomas},
  booktitle={Proceedings of The ACM Collective Intelligence Conference},
  pages={1--11},
  year={2023}
}

@inproceedings{wadinambiarachchi2024effects,
  title={The effects of generative AI on design fixation and divergent thinking},
  author={Wadinambiarachchi, Samangi and Kelly, Ryan M and Pareek, Saumya and Zhou, Qiushi and Velloso, Eduardo},
  booktitle={Proceedings of the 2024 CHI conference on human factors in computing systems},
  pages={1--18},
  year={2024}
}

@inproceedings{rafner2023picture,
  title={Picture this: AI-assisted image generation as a resource for problem construction in creative problem-solving},
  author={Rafner, Janet and Zana, Blanka and Dalsgaard, Peter and Biskjaer, Michael Mose and Sherson, Jacob},
  booktitle={Proceedings of the 15th Conference on Creativity and Cognition},
  pages={262--268},
  year={2023}
}

@article{shen2010online,
  title={An online visualization tool for Internet-based local townscape design},
  author={Shen, Zhenjiang and Kawakami, Mitsuhiko},
  journal={Computers, Environment and Urban Systems},
  volume={34},
  number={2},
  pages={104--116},
  year={2010},
  publisher={Elsevier}
}

@article{akbar2025use,
  title={The use of gamification for participatory smart city planning of Indonesia's new capital},
  author={Akbar, Poeti Nazura Gulfira and Maulidiyanti, Mareta and Wiwesa, Ngurah Rangga and Auliya, Anisatul},
  journal={International Journal of Urban Sciences},
  volume={29},
  number={sup1},
  pages={210--231},
  year={2025},
  publisher={Taylor \& Francis}
}

@article{paananen2024using,
  title={Using text-to-image generation for architectural design ideation},
  author={Paananen, Ville and Oppenlaender, Jonas and Visuri, Aku},
  journal={International Journal of Architectural Computing},
  volume={22},
  number={3},
  pages={458--474},
  year={2024},
  publisher={Sage publications Sage UK: London, England}
}

@article{cherry2014quantifying,
  title={Quantifying the creativity support of digital tools through the creativity support index},
  author={Cherry, Erin and Latulipe, Celine},
  journal={ACM Transactions on Computer-Human Interaction (TOCHI)},
  volume={21},
  number={4},
  pages={1--25},
  year={2014},
  publisher={ACM New York, NY, USA}
}

@article{guridi2025image,
  title     = {{Image Generative AI} to Design Public Spaces:
               A Reflection of How {AI} Could Improve Co-Design of Public Parks},
  author    = {Guridi, Jose A. and Cheyre, Cristobal and Goula, Maria
               and Santo, Duarte and Humphreys, Lee
               and Shankar, Aishwarya and Souras, Achilleas},
  journal   = {Digital Government: Research and Practice},
  volume    = {6},
  number    = {1},
  articleno = {7},
  numpages  = {14},
  year      = {2025},
  publisher = {Association for Computing Machinery},
  doi       = {10.1145/3656588}
}

@article{he2025recall,
  author    = {He, Zhiting and Su, Jiayi and Chen, Li
               and Wang, Tianqi and Lc, Ray},
  title     = {{'I Recall the Past'}: Exploring How People
               Collaborate with Generative {AI} to Create
               Cultural Heritage Narratives},
  journal   = {Proceedings of the ACM on Human-Computer Interaction},
  volume    = {9},
  number    = {2},
  articleno = {CSCW108},
  numpages  = {30},
  year      = {2025},
  month     = may,
  publisher = {Association for Computing Machinery},
  doi       = {10.1145/3711006}
}

@article{alkhateeb2024post,
  title={Post-conflict urban landscape storytelling: Two approaches to contemporary virtual visualisation of oral narratives},
  author={Alkhateeb, Ghieth and Storie, Joanna and K{\"u}lvik, Mart},
  journal={Land},
  volume={13},
  number={4},
  pages={406},
  year={2024},
  publisher={MDPI}
}

@inproceedings{hoggenmueller2023creative,
  title={Creative AI for HRI design explorations},
  author={Hoggenmueller, Marius and Lupetti, Maria Luce and Van Der Maden, Willem and Grace, Kazjon},
  booktitle={Companion of the 2023 ACM/IEEE International Conference on Human-Robot Interaction},
  pages={40--50},
  year={2023}
}

@inproceedings{von2024equipping,
  title={Equipping participation formats with generative AI: a case study predicting the future of a metropolitan city in the year 2040},
  author={von Brackel-Schmidt, Constantin and Ku{\v{c}}evi{\'c}, Emir and Leible, Stephan and Simic, Dejan and G{\"u}c{\"u}k, Gian-Luca and Schmidt, Felix N},
  booktitle={International Conference on Human-Computer Interaction},
  pages={270--285},
  year={2024},
  organization={Springer}
}

@inproceedings{verheijden2023collaborative,
  title={Collaborative diffusion: Boosting designerly co-creation with generative AI},
  author={Verheijden, Mathias Peter and Funk, Mathias},
  booktitle={Extended abstracts of the 2023 CHI conference on human factors in computing systems},
  pages={1--8},
  year={2023}
}

@INPROCEEDINGS{lin2024text,
  title     = "Text-to-image {AI} as a catalyst for semantic convergence in
               creative collaborations",
  author    = "Lin, Pei-Ying and Andersen, Kristina and Schmidt, Ralf and
               Schoenmakers, Sanne and Hofmeyer, Hèrm and Pauwels, Pieter and
               Ijsselsteijn, Wijnand",
  booktitle = "Designing Interactive Systems Conference",
  publisher = "ACM",
  pages     = "2753--2767",
  year      =  2024,
  doi       = "10.1145/3643834.3661543"
}

@ARTICLE{awashra2025generative,
  title     = "Generative {AI} text-to-image for community participation in
               landscape planning",
  author    = "Awashra, Ishraq and Thompson, Aaron W and Floress, Kristin and
               Arbuckle, J Gordon and Church, Sarah P and Genskow, Ken and
               Prokopy, Linda S and Rui, Yichao and Tesdell, Omar",
  journal   = "Landscape and Urban Planning",
  publisher = "Elsevier BV",
  volume    =  264,
  pages     =  105464,
  month     =  dec,
  year      =  2025,
  doi       = "10.1016/j.landurbplan.2025.105464",
  language  = "en"
}

@article{guridi2024fake,
  author    = {Guridi, Jose A. and Hwang, Angel Hsing-Chi
               and Santo, Duarte and Goula, Maria
               and Cheyre, Cristobal and Humphreys, Lee
               and Rangel, Marco},
  title     = {From Fake Perfects to Conversational Imperfects:
               Exploring Image-Generative {AI} as a Boundary
               Object for Participatory Design of Public Spaces},
  journal   = {Proceedings of the ACM on Human-Computer Interaction},
  volume    = {9},
  number    = {2},
  articleno = {CSCW014},
  numpages  = {33},
  year      = {2025},
  publisher = {Association for Computing Machinery},
  doi       = {10.1145/3710912}
}

@ARTICLE{pedersen2020staging,
  title     = "Staging negotiation spaces: A co-design framework",
  author    = "Pedersen, Signe",
  journal   = "Design Studies",
  publisher = "Elsevier BV",
  volume    =  68,
  pages     = "58--81",
  month     =  may,
  year      =  2020,
  doi       = "10.1016/j.destud.2020.02.002",
  language  = "en"
}

@incollection{vinck2012accessing,
  author    = {Vinck, Dominique},
  title     = {Accessing Material Culture by Following
               Intermediary Objects},
  booktitle = {An Ethnography of Global Landscapes and Corridors},
  editor    = {Naidoo, Loshini},
  publisher = {IntechOpen},
  pages     = {89--108},
  year      = {2012},
  doi       = {10.5772/34719}
}

@ARTICLE{sadorge2024generative,
  title     = "The generative role of objects in infrastructure design: A case
               of designing a system for continuity of care",
  author    = "Sadorge, Christopher and Nerland, Monika and Grisot, Miria",
  journal   = "Computer Supported Cooperative Work: CSCW: An International
               Journal",
  publisher = "Springer Science and Business Media LLC",
  volume    =  33,
  number    =  4,
  pages     = "1115--1149",
  month     =  dec,
  year      =  2024,
  doi       = "10.1007/s10606-023-09485-4",
  language  = "en"
}

@ARTICLE{pedersen2022navigating,
  title     = "Navigating collaborative open innovation projects: Staging
               negotiations of actors' concerns",
  author    = "Pedersen, Signe and Bogers, Marcel L A M and Clausen, Christian",
  journal   = "Creativity and Innovation Management",
  publisher = "Wiley",
  volume    =  31,
  number    =  2,
  pages     = "306--321",
  month     =  jun,
  year      =  2022,
  doi       = "10.1111/caim.12492",
  language  = "en"
}

@ARTICLE{pedersen2025staging,
  title     = "Staging situated negotiation games for (re)designing local
               healthcare facilities and services",
  author    = "Pedersen, Signe and Dorland, Jens",
  journal   = "Design for Health (Abingdon, England)",
  publisher = "Informa UK Limited",
  volume    =  9,
  number    =  1,
  pages     = "94--120",
  month     =  "2~" # jan,
  year      =  2025,
  doi       = "10.1080/24735132.2025.2454731",
  language  = "en"
}

@book{latour2005reassembling,
  author    = {Latour, Bruno},
  title     = {Reassembling the Social: An Introduction to Actor-Network-Theory},
  year      = {2005},
  publisher = {Oxford University Press},
  address   = {Oxford}
}

@article{storni2015designing,
  title     = "Designing things together: intersections of co-design and
               actor–network theory",
  author    = "Storni, Cristiano and Binder, Thomas and Linde, Per and Stuedahl,
               Dagny",
  journal   = "CoDesign",
  publisher = "Informa UK Limited",
  volume    =  11,
  number    = "3-4",
  pages     = "149--151",
  month     =  "2~" # oct,
  year      =  2015,
  doi       = "10.1080/15710882.2015.1081442"
}

@article{clausen2015social,
  title    = "From the social shaping of technology to the staging of temporary
              spaces of innovation: A case of participatory innovation",
  author   = "Clausen, Christian and Gunn, W",
  journal  = "Science \& Technology Studies",
  volume   =  28,
  number   =  1,
  pages    = "73--94",
  month    =  "15~" # apr,
  year     =  2015,
  language = "en"
}

@ARTICLE{fischer2021design,
  title     = "Design multiple: How different configurations of participation
               matter in design practice",
  author    = "Fischer, Björn and Östlund, Britt and Peine, Alexander",
  journal   = "Design Studies",
  publisher = "Elsevier BV",
  volume    =  74,
  pages     =  101016,
  month     =  may,
  year      =  2021,
  doi       = "10.1016/j.destud.2021.101016",
  language  = "en"
}

@INCOLLECTION{ehn2011participation,
  title     = "Participation in design things",
  author    = "Ehn, Pelle",
  booktitle = "Design Things",
  publisher = "The MIT Press",
  pages     = "157--182",
  month     =  "30~" # sep,
  year      =  2011,
  doi       = "10.7551/mitpress/8262.003.0011",
  language  = "en"
}

@INPROCEEDINGS{jenkins2016object,
  title     = "Object-Oriented Publics",
  author    = "Jenkins, Tom and Le Dantec, Christopher A and DiSalvo, Carl and
               Lodato, Thomas and Asad, Mariam",
  booktitle = "Proceedings of the 2016 CHI Conference on Human Factors in
               Computing Systems",
  publisher = "ACM",
  pages     = "827--839",
  month     =  "7~" # may,
  year      =  2016,
  doi       = "10.1145/2858036.2858565"
}

@ARTICLE{bodker2017tying,
  title     = "Tying knots: Participatory infrastructuring at work",
  author    = "Bødker, Susanne and Dindler, Christian and Iversen, Ole Sejer",
  journal   = "Computer Supported Cooperative Work: CSCW: An International
               Journal",
  publisher = "Springer Science and Business Media LLC",
  volume    =  26,
  number    = "1-2",
  pages     = "245--273",
  month     =  apr,
  year      =  2017,
  doi       = "10.1007/s10606-017-9268-y",
  language  = "en"
}

@incollection{vinck1994mediating,
  author    = {Vinck, Dominique and Jeantet, Alain},
  title     = {Mediating and Commissioning Objects in the
               Sociotechnical Process of Product Design:
               A Conceptual Approach},
  booktitle = {Management and New Technology:
               Designs, Networks and Strategies},
  editor    = {Maclean, Donald and Saviotti, Paolo
               and Vinck, Dominique},
  series    = {COST Social Science Series},
  volume    = {2},
  publisher = {EC Directorate General Science R\&D},
  address   = {Brussels},
  pages     = {111--129},
  year      = {1995}
}

@ARTICLE{boujut2003intermediary,
  title     = "Intermediary objects as a means to foster co-operation in
               engineering design",
  author    = "Boujut, Jean-François and Blanco, Eric",
  journal   = "Computer Supported Cooperative Work: CSCW: An International
               Journal",
  publisher = "Springer Science and Business Media LLC",
  volume    =  12,
  number    =  2,
  pages     = "205--219",
  month     =  jun,
  year      =  2003,
  doi       = "10.1023/a:1023980212097",
  language  = "en"
}

@ARTICLE{eckert2003role,
  title     = "The role of objects in design co-operation: Communication through
               physical or virtual objects",
  author    = "Eckert, Claudia and Boujut, Jean-François",
  journal   = "Computer Supported Cooperative Work: CSCW: An International
               Journal",
  publisher = "Springer Science and Business Media LLC",
  volume    =  12,
  number    =  2,
  pages     = "145--151",
  month     =  jun,
  year      =  2003,
  doi       = "10.1023/a:1023954726209",
  language  = "en"
}

@ARTICLE{paavola2019dynamics,
  title     = "Dynamics of design collaboration: {BIM} models as intermediary
               digital objects",
  author    = "Paavola, Sami and Miettinen, Reijo",
  journal   = "Computer Supported Cooperative Work: CSCW: An International
               Journal",
  publisher = "Springer Science and Business Media LLC",
  volume    =  28,
  number    = "1-2",
  pages     = "1--23",
  month     =  apr,
  year      =  2019,
  doi       = "10.1007/s10606-018-9306-4",
  language  = "en"
}

@article{wither2025using,
  title     = {Using Performative Objects to Foster Creativity
               in an Education Setting to Tackle Complex Challenges},
  author    = {Winther, Cathrine Marie Skovbo and Brodersen, S{\o}sser},
  journal   = {Design and Technology Education: An International Journal},
  volume    = {30},
  number    = {1},
  pages     = {26--44},
  year      = {2025},
  doi       = {10.24377/DTEIJ.article3041}
}

@techreport{quantum2007community,
  author      = {{Quantum Market Research}},
  title       = {Community Interest and Engagement with Science and Technology in Victoria: Research Report -- June 2007},
  institution = {{Victorian Department of Innovation, Industry and Regional Development}},
  address     = {Melbourne, Australia},
  year        = {2008},
  isbn        = {9781921572005}
}

@ARTICLE{stilgoe2013developing,
  title     = "Developing a framework for responsible innovation",
  author    = "Stilgoe, Jack and Owen, Richard and Macnaghten, Phil",
  journal   = "Research Policy",
  publisher = "Elsevier BV",
  volume    =  42,
  number    =  9,
  pages     = "1568--1580",
  month     =  nov,
  year      =  2013,
  doi       = "10.1016/j.respol.2013.05.008",
  language  = "en"
}

@article{braun2019reflecting,
  author    = {Braun, Virginia and Clarke, Victoria},
  title     = {Reflecting on Reflexive Thematic Analysis},
  journal   = {Qualitative Research in Sport, Exercise and Health},
  year      = {2019},
  volume    = {11},
  number    = {4},
  pages     = {589--597},
  publisher = {Routledge},
  doi       = {10.1080/2159676X.2019.1628806}
}

@book{braun2022thematic,
  author    = {Braun, Virginia and Clarke, Victoria},
  title     = {Thematic Analysis: A Practical Guide},
  publisher = {SAGE Publications},
  address   = {London},
  year      = {2022},
  isbn      = {9781473953246}
}

@INPROCEEDINGS{wong2020infrastructural,
  title     = "Infrastructural speculations: Tactics for designing and
               interrogating lifeworlds",
  author    = "Wong, Richmond Y and Khovanskaya, Vera and Fox, Sarah E and
               Merrill, Nick and Sengers, Phoebe",
  booktitle = "Proceedings of the 2020 CHI Conference on Human Factors in
               Computing Systems",
  publisher = "ACM",
  pages     = "1--15",
  month     =  apr,
  year      =  2020,
  doi       = "10.1145/3313831.3376515"
}

@ARTICLE{star1996steps,
  title     = "Steps toward an ecology of infrastructure: Design and access for
               large information spaces",
  author    = "Star, Susan Leigh and Ruhleder, Karen",
  editor    = "Yates, Joanne and VanMaanen, John",
  journal   = "Information Systems Research",
  publisher = "Institute for Operations Research and the Management Sciences
               (INFORMS)",
  volume    =  7,
  number    =  1,
  pages     = "111--134",
  month     =  mar,
  year      =  1996,
  doi       = "10.1287/isre.7.1.111",
  language  = "en"
}

@inproceedings{neumann1996making,
  author    = {Neumann, Laura J. and Star, Susan Leigh},
  title     = {Making Infrastructure:
               The Dream of a Common Language},
  booktitle = {Proceedings of the Participatory Design
               Conference (PDC '96)},
  editor    = {Blomberg, Janette and Kensing, Finn
               and Dykstra-Erickson, Elisabeth},
  publisher = {Computer Professionals for Social Responsibility},
  pages     = {231--240},
  year      = {1996},
  url       = {https://ojs.ruc.dk/index.php/pdc/article/view/153}
}

@incollection{selin2015against,
  title     = {Against Blank Slate Futuring:
               Noticing Obduracy in the City through Experiential
               Methods of Public Engagement},
  author    = {Selin, Cynthia and Sadowski, Jathan},
  booktitle = {Remaking Participation:
               Science, Environment and Emergent Publics},
  editor    = {Chilvers, Jason and Kearnes, Matthew},
  pages     = {218--237},
  year      = {2015},
  publisher = {Routledge},
  address   = {New York},
  doi       = {10.4324/9780203797693}
}

@ARTICLE{selin2017experiments,
  title     = "Experiments in engagement: Designing public engagement with
               science and technology for capacity building",
  author    = "Selin, Cynthia and Rawlings, Kelly Campbell and de
               Ridder-Vignone, Kathryn and Sadowski, Jathan and Altamirano
               Allende, Carlo and Gano, Gretchen and Davies, Sarah R and Guston,
               David H",
  journal   = "Public Understanding of Science (Bristol, England)",
  publisher = "SAGE Publications",
  volume    =  26,
  number    =  6,
  pages     = "634--649",
  month     =  aug,
  year      =  2017,
  doi       = "10.1177/0963662515620970",
  language  = "en"
}

@ARTICLE{andersen2015participation,
  title     = "Participation as a matter of concern in participatory design",
  author    = "Andersen, Lars Bo and Danholt, Peter and Halskov, Kim and Hansen,
               Nicolai Brodersen and Lauritsen, Peter",
  journal   = "CoDesign",
  publisher = "Informa UK Limited",
  volume    =  11,
  number    = "3-4",
  pages     = "250--261",
  month     =  oct,
  year      =  2015,
  doi       = "10.1080/15710882.2015.1081246",
  language  = "en"
}

@article{chilvers2020remaking,
  author  = {Jason Chilvers and Matthew Kearnes},
  title   = {Remaking Participation in Science and Democracy},
  journal = {Science, Technology, \& Human Values},
  volume  = {45},
  number  = {3},
  pages   = {347--380},
  year    = {2020},
  doi     = {10.1177/0162243919850885}
}

@inproceedings{vines2013configuring,
  author    = {Vines, John and Clarke, Rachel and Wright, Peter
               and McCarthy, John and Olivier, Patrick},
  title     = {Configuring Participation:
               On How We Involve People in Design},
  booktitle = {Proceedings of the SIGCHI Conference
               on Human Factors in Computing Systems},
  publisher = {Association for Computing Machinery},
  pages     = {429--438},
  year      = {2013},
  doi       = {10.1145/2470654.2470716}
}

@ARTICLE{bratteteig2016unpacking,
  title     = "Unpacking the notion of participation in participatory design",
  author    = "Bratteteig, Tone and Wagner, Ina",
  journal   = "Computer Supported Cooperative Work: CSCW: An International
               Journal",
  publisher = "Springer Science and Business Media LLC",
  volume    =  25,
  number    =  6,
  pages     = "425--475",
  month     =  dec,
  year      =  2016,
  doi       = "10.1007/s10606-016-9259-4",
  language  = "en"
}

@ARTICLE{davies2013rules,
  title     = "The rules of engagement: Power and interaction in dialogue events",
  author    = "Davies, Sarah R",
  journal   = "Public Understanding of Science (Bristol, England)",
  publisher = "SAGE Publications",
  volume    =  22,
  number    =  1,
  pages     = "65--79",
  month     =  jan,
  year      =  2013,
  doi       = "10.1177/0963662511399685",
  language  = "en"
}

@ARTICLE{felt2018imagine,
  title     = "{IMAGINE} {RRI}. A card-based method for reflecting on
               responsibility in life science research",
  author    = "Felt, Ulrike and Fochler, Maximilian and Sigl, Lisa",
  journal   = "Journal of Responsible Innovation",
  publisher = "Informa UK Limited",
  volume    =  5,
  number    =  2,
  pages     = "201--224",
  month     =  may,
  year      =  2018,
  doi       = "10.1080/23299460.2018.1457402",
  language  = "en"
}

@article{kubota2026nudging,
  author  = {Yuki Kubota and Kazushi Maruya and Maki Ogawa and
             Taiki Fukiage and Takumi Yokosaka and Yuri Mikawa},
  title   = {Public perceptions of infrastructural digital nudging:
             an experience-based study in a quasi-public setting},
  journal = {AI \& SOCIETY},
  year    = {2026},
  doi     = {10.1007/s00146-026-03370-6},
  url     = {https://doi.org/10.1007/s00146-026-03370-6}
}

@misc{mikawa2026visual,
  author        = {Yuri Mikawa and Taiki Fukiage and Yuki Kubota and
                   Takumi Yokosaka and Maki Ogawa and Kazushi Maruya},
  title         = {Visual-Motion-Induced Modulation of Pedestrian
                   Trajectories Using Spatially Distributed
                   Multi-Display Signage in Public Spaces},
  year          = {2026},
  eprint        = {2609.11088},
  archivePrefix = {arXiv},
  primaryClass  = {cs.HC},
  doi           = {10.48550/arXiv.2609.11088},
  url           = {https://arxiv.org/abs/2609.11088}
}

@incollection{akrich1992description,
  author    = {Akrich, Madeleine},
  title     = {The De-Scription of Technical Objects},
  booktitle = {Shaping Technology/Building Society:
               Studies in Sociotechnical Change},
  editor    = {Bijker, Wiebe E. and Law, John},
  publisher = {MIT Press},
  address   = {Cambridge, MA},
  pages     = {205--224},
  year      = {1992}
}

\newpage
\appendix
\section*{Appendix}

\counterwithin{figure}{section}
\counterwithin{table}{section}
\renewcommand{\thefigure}{\thesection\arabic{figure}}
\renewcommand{\thetable}{\thesection\arabic{table}}

\section{Procedure of Work 1}
\label{app:work1}
\begin{figure*}[htb]
  \centering
  \includegraphics[keepaspectratio,width=1.00\linewidth]{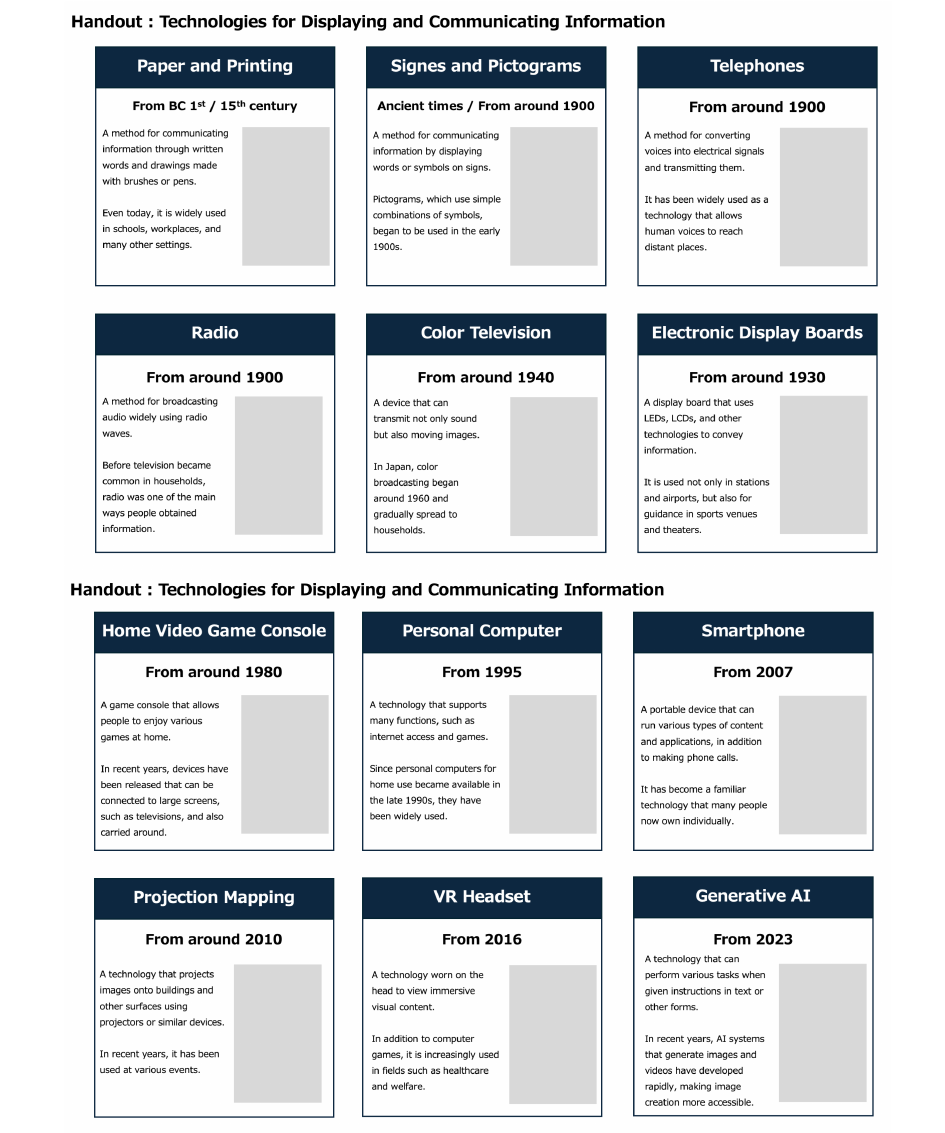}
  \caption{English reproduction of the handout used in Work 1, introducing twelve information-display and communication technologies. Photographs included in the original handout were omitted because of copyright restrictions on web-sourced images.}
  \label{fig:procedures-ws1}
\end{figure*}

After briefly introducing themselves within their groups, participants first received the handout (Figure~\ref{fig:procedures-ws1}), which presented a range of familiar information-display and communication technologies. Each participant selected one technology to compare with digital signage. When multiple participants in the same group chose the same technology, they were encouraged to select different ones so that the group could consider a broader range of comparisons. Participants then wrote similarities and differences between digital signage and their selected technology on sticky notes and placed them on an A4 sheet corresponding to that technology.

After approximately three minutes of individual reflection, participants passed their sheets to the person on their left and added further observations to those prepared by other group members. This rotation continued across the group, allowing participants to compare digital signage with multiple familiar technologies and consider how its distinctive characteristics might be used. This activity encouraged participants to consider digital signage in relation to familiar technologies before envisioning its potential uses in Work 2.

\section{Questionnaire contents and analysis}
\label{app:questionnaire}

\subsection{Questionnaire overview}
Participants completed a questionnaire after the workshop. The full set of items is provided in Appendix~\ref{app:questionnaire-full}. The questionnaire comprised four components: (1) demographic information; (2) interest in science and technology and prior experience with generative AI; (3) perceptions of creative engagement supported by the workshop materials and activities, informed by the Creativity Support Index (CSI)~\cite{cherry2014quantifying}; and (4) broader evaluations of the workshop informed by the responsible research and innovation (RRI) framework~\cite{stilgoe2013developing}.

Participants first reported their age group and gender. Interest in science and technology was assessed using an item adapted from a previous study~\cite{quantum2007community}: ``Are you interested in science and technology?'' Prior experience with generative AI was assessed by asking participants to report their frequency of use.

To examine creative engagement, we adapted selected dimensions of the CSI, which comprises six dimensions: collaboration, enjoyment, exploration, expressiveness, immersion, and results worth effort~\cite{cherry2014quantifying}. The CSI has previously been adapted to evaluate generative AI in architectural design~\cite{paananen2024using}. Because our workshop involved multiple materials and activities that could support creativity in different ways, we developed 14 items informed by selected CSI dimensions to assess participants' perceptions of specific workshop activities. Two additional items assessed how closely the initial image prepared before group discussion and the final image developed through successive revisions matched each participant's imagined future.

We also developed six items informed by the four dimensions of RRI---anticipation, reflexivity, inclusion, and responsiveness~\cite{stilgoe2013developing}. These items assessed whether the workshop encouraged participants to broaden their thinking about future technologies and everyday life, reflect on their implications, engage with diverse perspectives, and connect the discussion to their own future actions and experiences.

\subsection{Full list of questionnaire items}
\label{app:questionnaire-full}
The post-workshop questionnaire consisted of demographic and background questions, 22 rating items, and one open-ended question. Unless otherwise noted, the rating items were answered on a seven-point scale ranging from $-3$ (Strongly disagree) to $+3$ (Strongly agree). The items below are English translations of the Japanese questionnaire administered to participants.

\textbf{Demographic and background information}
\begin{itemize}
    \item Gender: Male / Female / Other
    \item Age: 10--19 / 20--29 / 30--39 / 40--49 / 50--59 / 60 or older
    \item Interest in science and technology: Very interested / Interested / Neither interested nor uninterested / Not interested / Not at all interested / Not sure
    \item Frequency of generative AI use: Several times a day or more / About once every 1--2 days / About once every 1--2 weeks / About once every 1--2 months / Have used it a few times / Have never used it
\end{itemize}

\textbf{Demonstration experience}
\begin{enumerate}
    \item The experience of walking between the large displays was enjoyable.
    \item I was highly engaged in the experience of walking between the large displays.
    \item The walking experience helped me clarify or broaden my thinking during the workshop.
\end{enumerate}

\textbf{Work 1: Comparing digital signage with other technologies}
\begin{enumerate}
    \setcounter{enumi}{3}
    \item Comparing digital signage with other technologies was enjoyable.
    \item Comparing digital signage with other technologies helped me clarify or broaden my thinking.
    \item Through comparison with other technologies, I was able to develop ideas or thoughts that I was satisfied with.
\end{enumerate}

\textbf{Work 2: Imagining a future city}
\begin{enumerate}
    \setcounter{enumi}{6}
    \item The materials containing photographs of present-day urban environments helped me clarify or broaden my thinking.
    \item The worksheet for analyzing the future city helped me clarify or broaden my thinking.
    \item Using the worksheet helped me develop ideas or thoughts that I was satisfied with.
    \item Hearing other participants' visions of the future city helped deepen the group's thinking.
    \item Using generative AI to visualize my own and others' ideas helped us share ideas and collaborate with one another.
    \item Using generative AI to visualize my own and others' ideas helped me clarify or broaden my thinking.
    \item The initial AI-generated image shown at the beginning of the group discussion was close to the future I had imagined.
    \item The AI-generated image completed through the group discussion was close to the future I had imagined.
\end{enumerate}

\textbf{Cross-group sharing}
\begin{enumerate}
    \setcounter{enumi}{14}
    \item Hearing other groups' visions of the future city broadened my own thinking.
    \item Using generative AI to visualize other groups' ideas helped deepen my thinking across the workshop as a whole.
\end{enumerate}

\textbf{Overall reflection on the workshop}
\begin{enumerate}
    \setcounter{enumi}{16}
    \item Through today's demonstration and workshop, I broadened my imagination of future technologies.
    \item Through today's demonstration and workshop, I broadened my imagination of future everyday life.
    \item Today's demonstration and workshop gave me an opportunity to think about how new technologies may affect society and everyday life.
    \item Today's demonstration and workshop gave me an opportunity to reconsider my own views.
    \item During today's demonstration and workshop, I felt comfortable expressing my own opinions.
    \item I think I can apply what I experienced in today's demonstration and workshop to my future life or actions.
\end{enumerate}

\textbf{Open-ended question}
\begin{enumerate}
    \setcounter{enumi}{22}
    \item Please freely describe anything you felt or thought about through the demonstration and workshop.
\end{enumerate}

\section{Questionnaire results: Participants' perceptions of the workshop}
\label{app:questionnaire-results}
\begin{figure}[htb]
  \centering
  \includegraphics[keepaspectratio,width=1.00\linewidth]{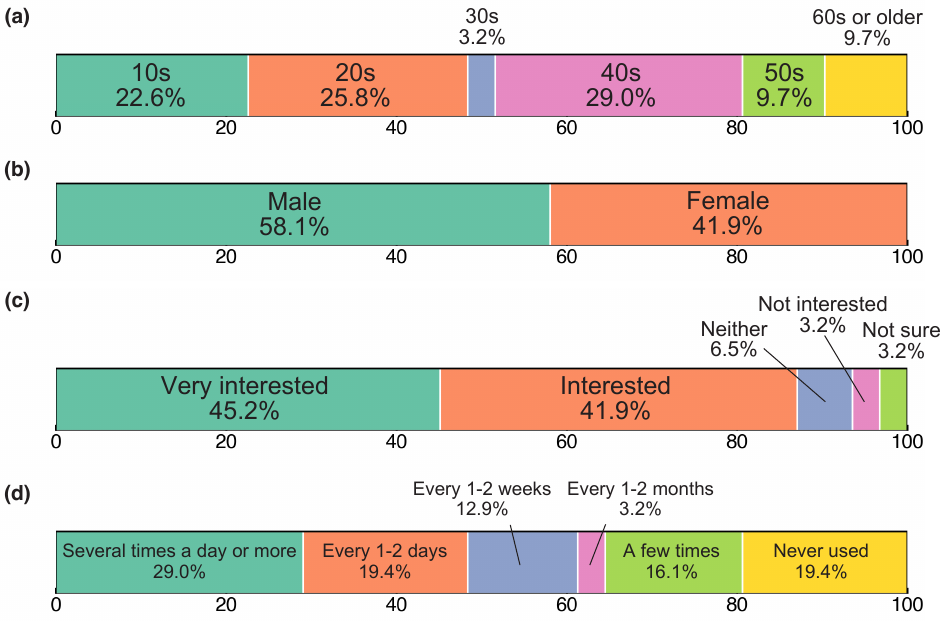}
  \caption{Participant demographics and background characteristics. (a) Age distribution. (b) Gender distribution. (c) Interest in science and technology. (d) Frequency of generative AI use.}
  \label{fig:results-profile}
\end{figure}

\begin{figure}[htb]
  \centering
  \includegraphics[keepaspectratio,width=1.00\linewidth]{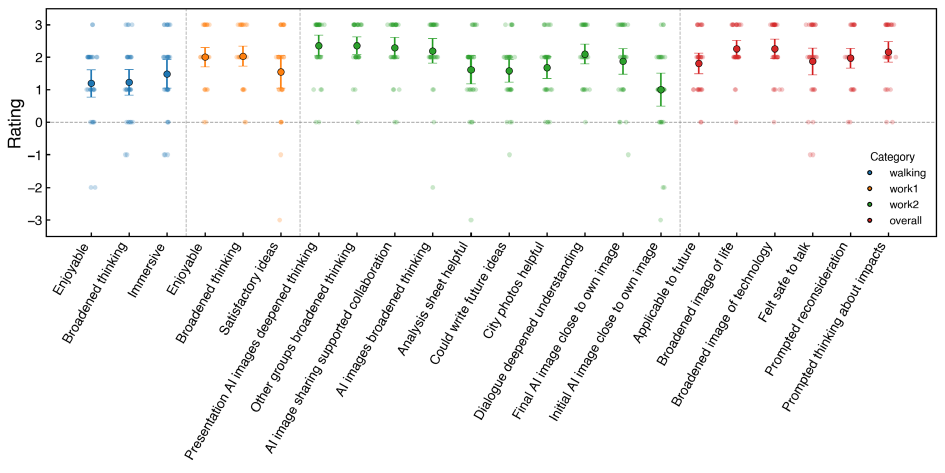}
  \caption{Participants' ratings of 22 post-workshop items ($N = 31$) on a seven-point scale ranging from $-3$ to $+3$. Small points indicate individual responses, while larger circles and error bars represent item means and 95\% confidence intervals. Items are grouped by workshop phase.}
  \label{fig:results-questionnaire}
\end{figure}

We examined participants' retrospective perceptions of the workshop using the post-workshop questionnaire. Figure~\ref{fig:results-profile} summarizes participants' demographic and background characteristics, and Figure~\ref{fig:results-questionnaire} presents individual responses together with mean ratings and a 95\% confidence interval for each questionnaire item. For presentation, the 22 rating items were grouped by workshop phase: demonstrations (Items 1–3), Work 1 (Items 4–6), Work 2 including cross-group sharing (Items 7–16), and overall reflection (Items 17–22).

Overall, mean ratings for all evaluated workshop activities were above the neutral point of the seven-point Likert scale. Items concerning generative AI also received positive ratings. Participants reported that visualizing their own and others' ideas with generative AI supported idea sharing and collaboration (Item 11; $M=2.29$, 95\% CI $[1.97, 2.61]$) and helped clarify or broaden their thinking (Item 12; $M=2.19$, 95\% CI $[1.81, 2.57]$). Viewing AI-generated images of other groups' ideas was also perceived to deepen their thinking (Item 16; $M=2.35$, 95\% CI $[2.03, 2.68]$). In addition, the final images developed through group discussion received higher ratings for matching participants' imagined futures (Item 14; $M=1.87$, 95\% CI $[1.48, 2.26]$) than the initial images presented before participants' ideas were incorporated (Item 13; $M=1.00$, 95\% CI $[0.49, 1.51]$).

Other workshop activities were also evaluated positively. Participants reported enjoyment (Item 1; $M=1.19$, 95\% CI $[0.77, 1.61]$) and engagement (Item 2; $M=1.48$, 95\% CI $[1.05, 1.92]$) during the walking demonstration. Photographs of present-day urban environments (Item 7; $M=1.68$, 95\% CI $[1.35, 2.01]$) and the worksheet (Item 8; $M = 1.61$, 95\% CI $[1.18, 2.05]$) were also perceived as helpful for clarifying or broadening participants' ideas. Interpersonal activities received positive ratings, including hearing other group members' ideas (Item 10; $M=2.10$, 95\% CI $[1.79, 2.40]$) and sharing visions across groups (Item 15; $M=2.35$, 95\% CI $[2.09, 2.62]$). Overall, participants reported that the workshop broadened their imagination of future technologies (Item 17; $M=2.26$, 95\% CI $[1.96, 2.56]$) and future everyday life (Item 18; $M= 2.26$, 95\% CI $[2.00, 2.52]$).

Taken together, the questionnaire results indicate that participants generally perceived generative AI as a useful resource for developing ideas and supporting collaboration within the workshop. They also perceived the images produced through discussion as more closely aligned with their imagined futures than the initial images. These findings, however, reflect participants' retrospective evaluations and do not establish how AI-generated images shaped the course of discussion or how particular concerns changed over time. The questionnaire therefore complements the process-oriented qualitative analysis in the main text.

\FloatBarrier
\newpage
\section{Additional results of the qualitative analysis}
\subsection{Coding system constructed through qualitative analysis}
\label{app:coding-system}
The following tables summarize the coding system developed through the qualitative analysis described in Section~\ref{sec:data-collection}. The codes are organized into three categories: \textit{structure} (workshop stages), \textit{content} (discussion topics and concerns), and \textit{performance} (how participants and facilitators articulated, interpreted, and negotiated ideas).

\begin{longtable}{@{}p{0.45\textwidth}p{0.55\textwidth}@{}}
\caption{Code names and definitions for workshop stages (Structure). Image generation and subsequent discussion were coded separately for each round. The number of rounds varied across groups depending on the number of image updates.}
\label{tab:codebook-structure}\\
\toprule
\textbf{Code} & \textbf{Definition} \\
\midrule
\endfirsthead
\toprule
\textbf{Code} & \textbf{Definition} \\
\midrule
\endhead
\midrule
\multicolumn{2}{r}{\footnotesize Continued on next page}\\
\endfoot
\bottomrule
\endlastfoot
Introduction and task briefing & Introduction and explanation of workshop tasks. \\
Initial sharing of ideas & Sharing ideas before image generation. \\
Image generation (Round 1) & Generation of the first group-specific AI image. \\
Discussion and image revision (Round 1) & Viewing the first image and discussing revisions. \\
Image generation (Round 2) & Generation of the second group-specific AI image. \\
Discussion and image revision (Round 2) & Viewing the second image and discussing revisions. \\
Image generation (Round 3) & Generation of the third group-specific AI image. \\
Discussion and image revision (Round 3) & Viewing the third image and discussing revisions. \\
Image generation (Round 4) & Generation of the fourth group-specific AI image. \\
Discussion and image revision (Round 4) & Viewing the fourth image and discussing revisions. \\
Image selection & Selection of a final image for presentation. \\
Summary and wrap-up & Summary and closing discussion. \\
\end{longtable}

\begin{longtable}{@{}p{0.35\textwidth}p{0.65\textwidth}@{}}
\caption{Code names and definitions for discussion topics and concerns (Content). Codes were not mutually exclusive; multiple codes could be assigned to the same segment when it reflected more than one aspect of the interaction.}\label{tab:codebook-content}\\
\toprule
\textbf{Code} & \textbf{Definition} \\
\midrule
\endfirsthead
\toprule
\textbf{Code} & \textbf{Definition} \\
\midrule
\endhead
\midrule
\multicolumn{2}{r}{\footnotesize Continued on next page}\\
\endfoot
\bottomrule
\endlastfoot
\multicolumn{2}{@{}l}{\textit{Values (aspirations and concerns)}} \\*
Maintaining daily life and health & References to the maintenance of everyday lives or health. \\
Inclusion and accessibility & References to diverse users, such as people with disabilities, socially vulnerable people, or foreign nationals, and the need to accommodate them. \\
Human-centered information and spatial design & Proposals for technology or spatial design that benefits people by considering their physical or cognitive characteristics. \\
Efficient and digitized future lifestyles & Descriptions of future daily life characterized by efficiency or automation. \\
Restorative and relaxing public spaces & References to public spaces that provide relaxation or a restorative experience. \\
Appropriate amount of information & References to the need to regulate the amount of information. \\
Harmony with nature and the landscape & References to the integration of digital technologies, including signage, with natural surroundings or the landscape. \\
Real-time information provision & References to the real-time presentation or updating of information. \\
Personalized and context-aware information provision & References to information tailored to individual needs, user characteristics, situations, or contexts, including translation and accommodations for diversity. \\
Power and cost requirements & References to electricity or financial costs required by the proposed technologies, including on-site power generation. \\
Secondary effects and social concerns & References to secondary effects, unintended consequences, or social concerns associated with an envisioned future. \\
\addlinespace[0.2em]
\multicolumn{2}{@{}l}{\textit{Technological and spatial characteristics}} \\*
Use of nonvisual modalities (e.g., sound or scent) & References to the need for or use of audio, sound, or other nonvisual modalities. \\
Intuitive, nonverbal, and multilingual information provision & References to information provision using pictograms or other nonverbal means, translation, or multiple languages. \\
Visual environmental features (e.g., light and color) & References to adjustments in lighting, color, or other properties of the visual environment. \\
People, places, and objects in future cities & References to components of imagined future cities, such as buildings or people. \\
Signage form and display configuration & References to the shape or display configuration of signage, such as transparent or small-scale displays. \\
Expanded signage locations and display surfaces & References to expanding the objects or places on which signage can be installed or information displayed. \\
Reconfiguration of mobility & References to changes in mobility, such as autonomous driving, separation of pedestrians and vehicles, or movable benches. \\
Floating signage and flying vehicles & References to floating signage and mobility, including flying cars. \\
\addlinespace[0.2em]
\multicolumn{2}{@{}l}{\textit{Purposes of use}} \\*
Leisure, tourism, and entertainment & References to uses for leisure, tourism, or entertainment. \\
Advertising and commercial applications & References to signage as a medium for commercial information or advertising distributed by particular businesses or individuals. \\
Traffic guidance and congestion relief & References to uses intended to guide pedestrian or vehicular traffic or reduce congestion. \\
Warnings and safety & References to uses intended to prevent accidents or alert people to risks. \\
Everyday information and guidance & References to providing necessary information or guidance in everyday situations. \\
Disaster information and guidance & References to providing information, guidance, or assistance in disaster situations. \\
\multicolumn{2}{@{}l}{\textit{Unclassified}} \\*
\end{longtable}

\begin{longtable}{@{}>{\raggedright\arraybackslash}p{0.35\textwidth}>{\raggedright\arraybackslash}p{0.65\textwidth}@{}}
\caption{Code names and definitions for the articulation of future visions, engagement with AI-generated images, and coordination of discussions (Performance). Codes were not mutually exclusive; multiple codes could be assigned to the same segment when it reflected more than one aspect of the interaction.}\label{tab:codebook-performance}\\
\toprule
\textbf{Code} & \textbf{Definition} \\
\midrule
\endfirsthead
\toprule
\textbf{Code} & \textbf{Definition} \\
\midrule
\endhead
\midrule
\multicolumn{2}{r}{\footnotesize Continued on next page}\\
\endfoot
\bottomrule
\endlastfoot
\multicolumn{2}{@{}l}{\textit{Ways of describing future visions}} \\*
Rejection of or distance from future visions & Expressions of rejection, discomfort, or distance toward a proposed future vision or generated image. \\
Futures different from present life & Descriptions of futures explicitly contrasted with present technologies, living spaces, or everyday life. \\
Futures continuous with present life & Descriptions of futures resembling or extending current lifestyles or technologies. \\
Knowledge-based future visions & Descriptions of envisioned futures grounded in existing technologies or specialized knowledge. \\
Future visions based on personal experience & Descriptions of envisioned futures drawing on participants' personal experiences. \\
Future visions based on concrete situations & Descriptions of envisioned futures referring to specific present-day situations or personal experiences. \\
Future visions based on assumptions or hypotheses & Descriptions of envisioned futures based on hypothetical conditions or assumptions. \\
Future visions based on technological features & Descriptions of envisioned futures referring to characteristics of technologies such as digital signage. \\
\addlinespace[0.2em]
\multicolumn{2}{@{}l}{\textit{Engagement with AI images and materials}} \\*
References to earlier materials & References to materials or discussions from earlier stages other than AI-generated images. \\
Interpretation of AI-generated images & Interpretations by participants or facilitators of elements or content depicted in AI-generated images. \\
Anthropomorphizing AI & Descriptions of AI in human-like terms by participants or facilitators. \\
Positive responses to or acceptance of AI and its images & Expressions of positive reactions to or acceptance of AI or AI-generated images. \\
Recognition of ideas reflected in AI images & Identification of where participants' own ideas or those of others are represented in AI-generated images. \\
Attribution of added or extrapolated content to AI & Statements that AI independently generated, supplemented, or extrapolated elements included in an image. \\
Discrepancies between AI images and participants' ideas & References to mismatches between what an AI-generated image depicts and the speaker's own ideas. \\
Limitations of AI outputs & References to limitations in the outputs or expressive capabilities of AI or AI-generated images. \\
\addlinespace[0.2em]
\multicolumn{2}{@{}l}{\textit{Discussion among participants}} \\*
Collaborative development of ideas & Instances in which multiple participants discuss the same topic and collectively develop the discussion. \\
Reframing and negotiation of discussion topics & Instances in which participants or facilitators reorganize issues, negotiate content, or introduce alternative perspectives. \\
Connections and differences between ideas & References to relationships or differences between the speaker's ideas and others' ideas, or among multiple ideas. \\
Clarification and elaboration of ideas & Exchanges in which participants or facilitators clarify or probe the meaning or background of statements. \\
Acknowledgment and acceptance of contributions & Instances in which a participant or facilitator affirms or accepts another participant's statement. \\
New insights and shifts in perspective & Explicit expressions of new insights or changes in perspective prompted by another statement. \\
Disclosure of knowledge and expertise & Disclosure of technical knowledge or expertise, excluding passages directly used to describe a knowledge-based future vision. \\
Reflections on the experience and discussion & Reflections on the experience of discussing ideas with the support of AI. \\
\addlinespace[0.2em]
\multicolumn{2}{@{}l}{\textit{Facilitator coordination}} \\*
Clarification of facilitator roles & Instances in which facilitators define or clarify their roles in discussions with participants. \\
Supporting ease of participation & Proposals for ground rules or discussion practices that make participation easier. \\
Managing idea sharing and speaking turns & Instances in which facilitators encourage sharing or manage speaking opportunities and turn-taking. \\
\end{longtable}

\newpage
\subsection{Discussion trajectories across the workshop groups}
\label{app:workshop-trajectories}

This section provides case-level accounts of how concerns and interactions developed across the eight workshop groups. Table~\ref{tab:theme3-cross-case} summarizes concerns that were sustained, rearticulated, or received less attention.  Figures~\ref{fig:results-3-chart0906A}--\ref{fig:results-3-chart0907D} trace selected developments involving AI-generated images, participants' responses, and other forms of mediation. The diagrams highlight key stages and the indicated links represent interpretive associations. T1--T4 refer to the corresponding analytical themes (Themes 1 to 4) in Section~\ref{sec:results}.

\begin{table}[htbp]
\caption{Cross-case overview of evolving concerns during image-based discussions (Theme 3). ``Sustained / rearticulated'' includes concerns elaborated or reconsidered through different proposals. ``Less-sustained'' indicates limited further exploration. ``New / prominent'' includes topics that newly emerged during subsequent discussion.}
\label{tab:theme3-cross-case}
\centering
\small
\begin{tabular}{l p{0.30\linewidth} p{0.30\linewidth} p{0.30\linewidth}}
\toprule
\textbf{Group} &
\textbf{Sustained / rearticulated} &
\textbf{Less sustained} &
\textbf{New / prominent} \\
\midrule

9/6-A &
Traffic / lane guidance;
Mobility reorganization;
Context-aware / real-time information &
Disaster-related information;
Everyday information;
Power requirements;
Park guidance &
Pedestrian--vehicle separation;
Aerial displays / flying vehicles \\
\addlinespace

9/6-B &
Human-friendly information design;
Expanded installation sites $\rightarrow$ small signage for narrow streets &
Advertising / guidance;
Traffic guidance / congestion relief;
Switchable displays &
Road / sidewalk widths;
Small signage / new mobility;
Visual environment \\
\addlinespace

9/6-C &
Safety warnings;
Context-aware information $\rightarrow$ personal / contextual information &
Everyday information / guidance;
Audio and Japanese-text guidance &
Appropriate information density;
Surveillance risks;
Landscape harmony \\
\addlinespace

9/6-D &
Expanded installation sites $\rightarrow$ visual environment;
Human-friendly information design &
Advertising / safety alerts;
Citizen-created video content;
Future with little need to go out &
Landscape considerations;
Personalized information \\
\addlinespace

9/7-A &
Projection onto trees / leaves $\rightarrow$ tree-integrated signage;
Relaxing public space;
Audio uses &
Advertising / everyday information;
Disaster / tourist guidance;
Cost requirements &
Display shapes;
Shade from large displays;
On-site solar power generation \\
\addlinespace

9/7-B &
Expanded display locations;
Everyday information / guidance $\rightarrow$ clear wayfinding / robots &
Waiting-time entertainment;
Translation / multilingual information &
Display size / shape adjustments;
Glare / power-use concerns;
Motion plausibility \\
\addlinespace

9/7-C &
Appropriate information density;
Flexible display sizes $\rightarrow$ AR-view information;
Context-aware shop guidance $\rightarrow$ personalized information;
Landscape harmony &
Everyday guidance / advertising; 
Audio information;
Multilingual / pictogram information;
Care for surroundings (noise); &
Mobility reorganization;
Flying signage / drones;
Pedestrian--vehicle separation \\
\addlinespace

9/7-D &
Digitized street instruments $\rightarrow$ expanded installation sites;
Compact holographic displays $\rightarrow$ semi-transparent / transparent displays &
Diverse mobility;
Smooth traffic;
Voice-based wayfinding; &
Landscape harmony;
On-site power generation \\
\bottomrule
\end{tabular}
\end{table}

\newpage
\subsubsection*{Group 9/6-A}
\begin{figure}[htb]
  \centering
  \includegraphics[keepaspectratio,width=1.0\linewidth]{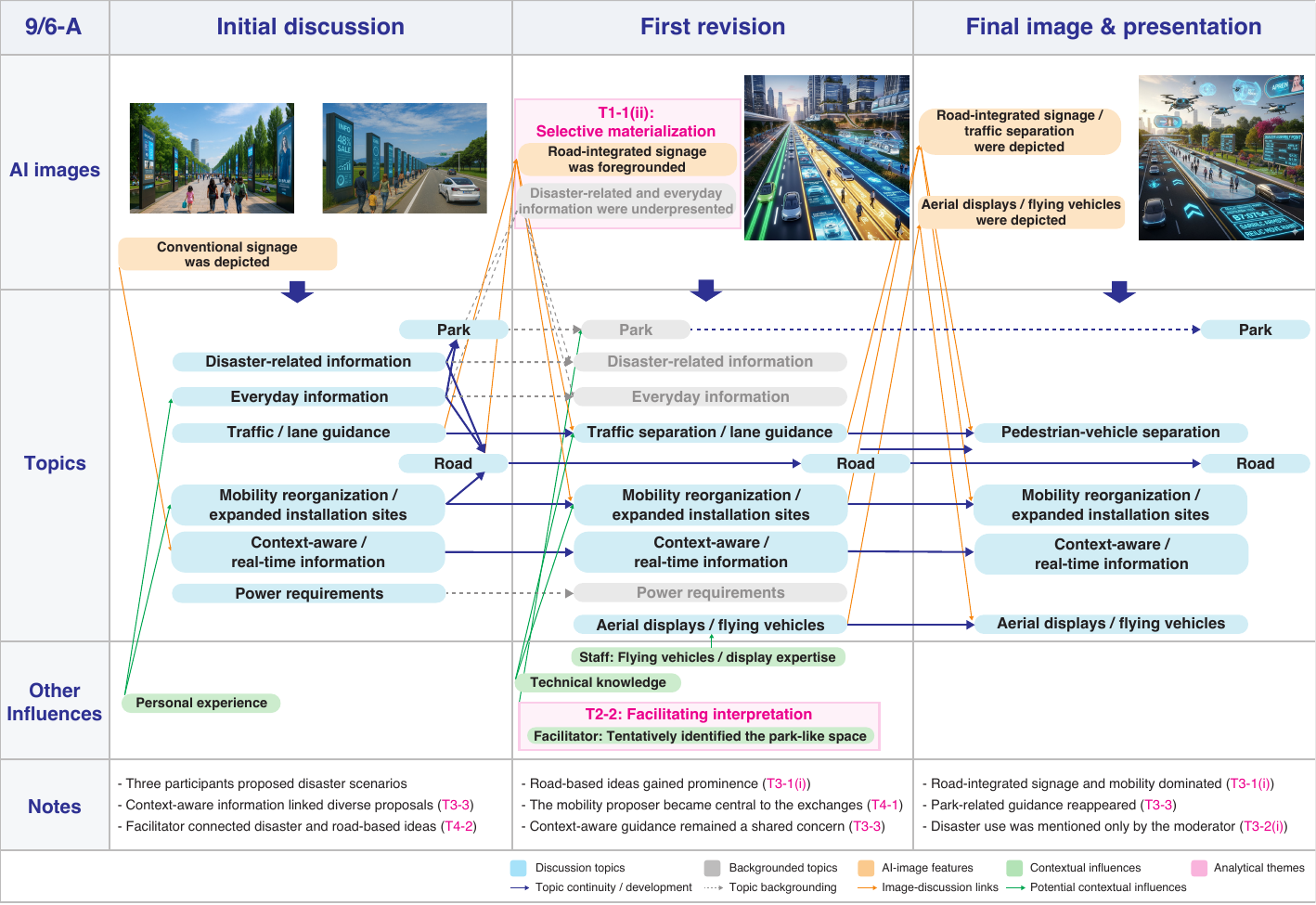}
  \caption{Evolution of topics and concerns in Group 9/6-A. The diagram traces the progression from initial discussion through image revision to the final presentation. Blue arrows indicate topic continuity or development, gray dashed arrows indicate backgrounding, orange arrows indicate links between AI images and discussion, and green arrows indicate possible influences from other resources.}
  \label{fig:results-3-chart0906A}
\end{figure}
Group 9/6-A initially discussed disaster-related information, everyday information, and traffic guidance, linked by a shared interest in context-aware information provision (Figure~\ref{fig:results-3-chart0906A}). Although disaster-related proposals were included in the prompt, the first revised image foregrounded road-integrated signage while underrepresenting disaster-related and everyday information (Theme 1-1(ii)). When a participant questioned where the park was depicted, the facilitator tentatively identified a park-like space (Theme 2-2). Subsequent discussion focused on traffic separation, lane guidance, and mobility reorganization (Theme 3-1(i)), drawing partly on participants' experiences and technical expertise from workshop staff (Theme 4-3). The participant who proposed road-integrated signage became central to these exchanges, whereas the participant who questioned the missing park made no further recorded contributions (Theme 4-1). Although specific disaster-related uses received little further attention without being explicitly rejected (Theme 3-2(i)), the broader concern with context-aware information persisted through proposals for real-time traffic guidance (Theme 3-3). The final image and presentation emphasized road-integrated signage and mobility, while park-related guidance briefly reappeared and disaster-related uses were mentioned only by the moderator.

\newpage
\subsubsection*{Group 9/6-B}
\begin{figure}[htb]
  \centering
  \includegraphics[keepaspectratio,width=1.0\linewidth]{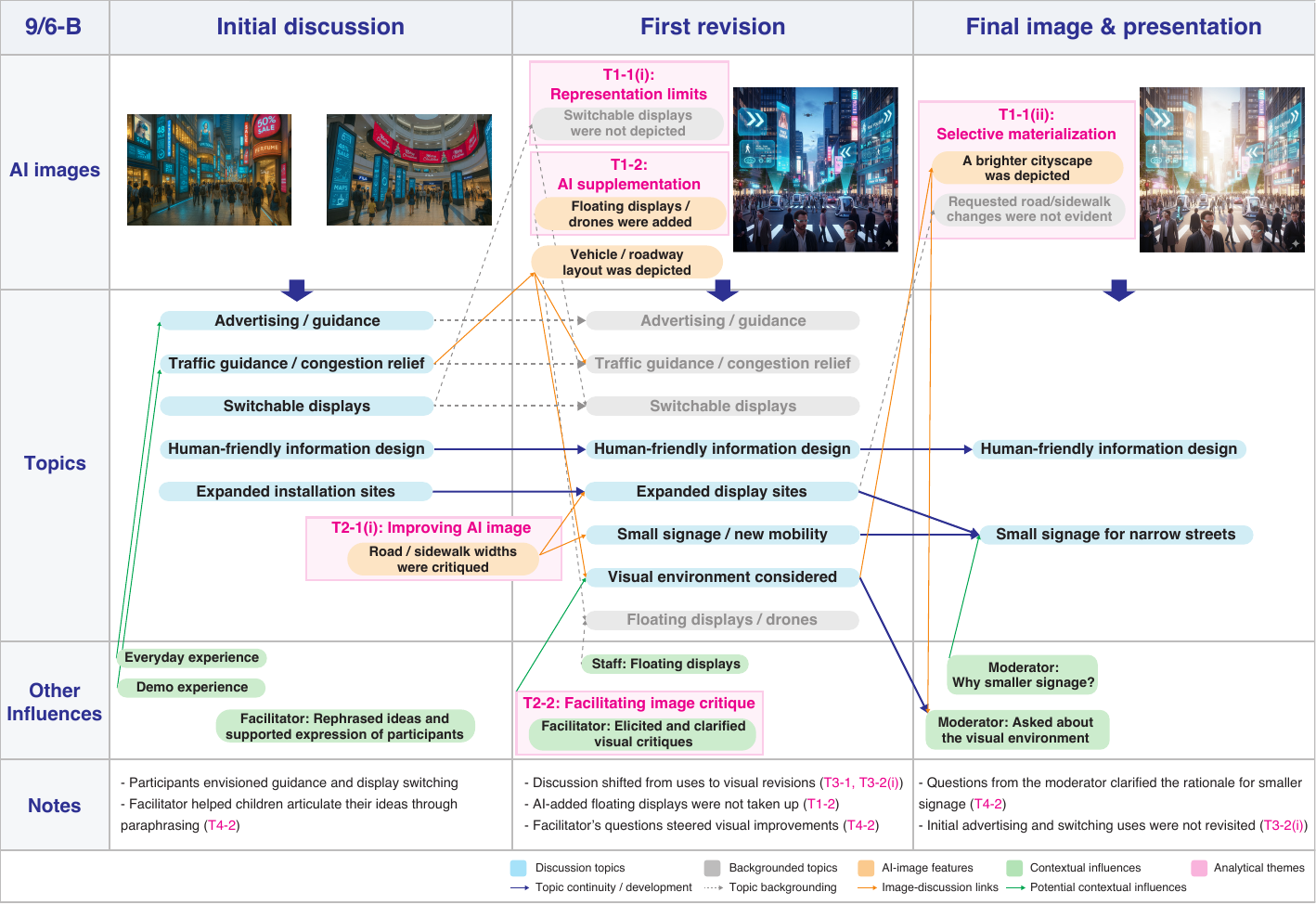}
  \caption{Evolution of topics and concerns in Group 9/6-B. Arrow and color conventions follow those in Figure~\ref{fig:results-3-chart0906A}.}
  \label{fig:results-3-chart0906B}
\end{figure}

Group 9/6-B initially proposed advertising, traffic guidance and congestion relief, switchable displays, and expanded installation sites (Figure~\ref{fig:results-3-chart0906B}). An earlier workshop demonstration also inspired a proposal for guiding pedestrian movement through psychological responses (Theme 4-3), while the facilitator helped the children articulate their ideas through paraphrasing (Theme 4-2). The first revised image depicted vehicle and roadway layouts but did not show switchable displays (Theme 1-1(i)). It also  introduced floating displays and drones during image generation, which received little attention (Theme 1-2). Encouraged by the facilitator's questions (Theme 2-2), participants critiqued road and sidewalk widths and proposed smaller signage and changes to the visual environment (Theme 2-1(i)). Discussion thus shifted toward spatial and visual design (Theme 3-1(ii)), while initial applications such as advertising, congestion relief, and display switching received less attention (Theme 3-2(i)). Nevertheless, the concern with human-friendly information design persisted through proposals for smaller signage on narrow streets. Although the final image depicted a brighter cityscape, the requested road and sidewalk revisions were not evident (Theme 1-1(ii)). During the final presentation, the moderator's questions helped clarify practical reasons for smaller signage, including installation and readability (Theme 4-2).

\newpage
\subsubsection*{Group 9/6-C}
\begin{figure}[htb]
  \centering
  \includegraphics[keepaspectratio,width=1.0\linewidth]{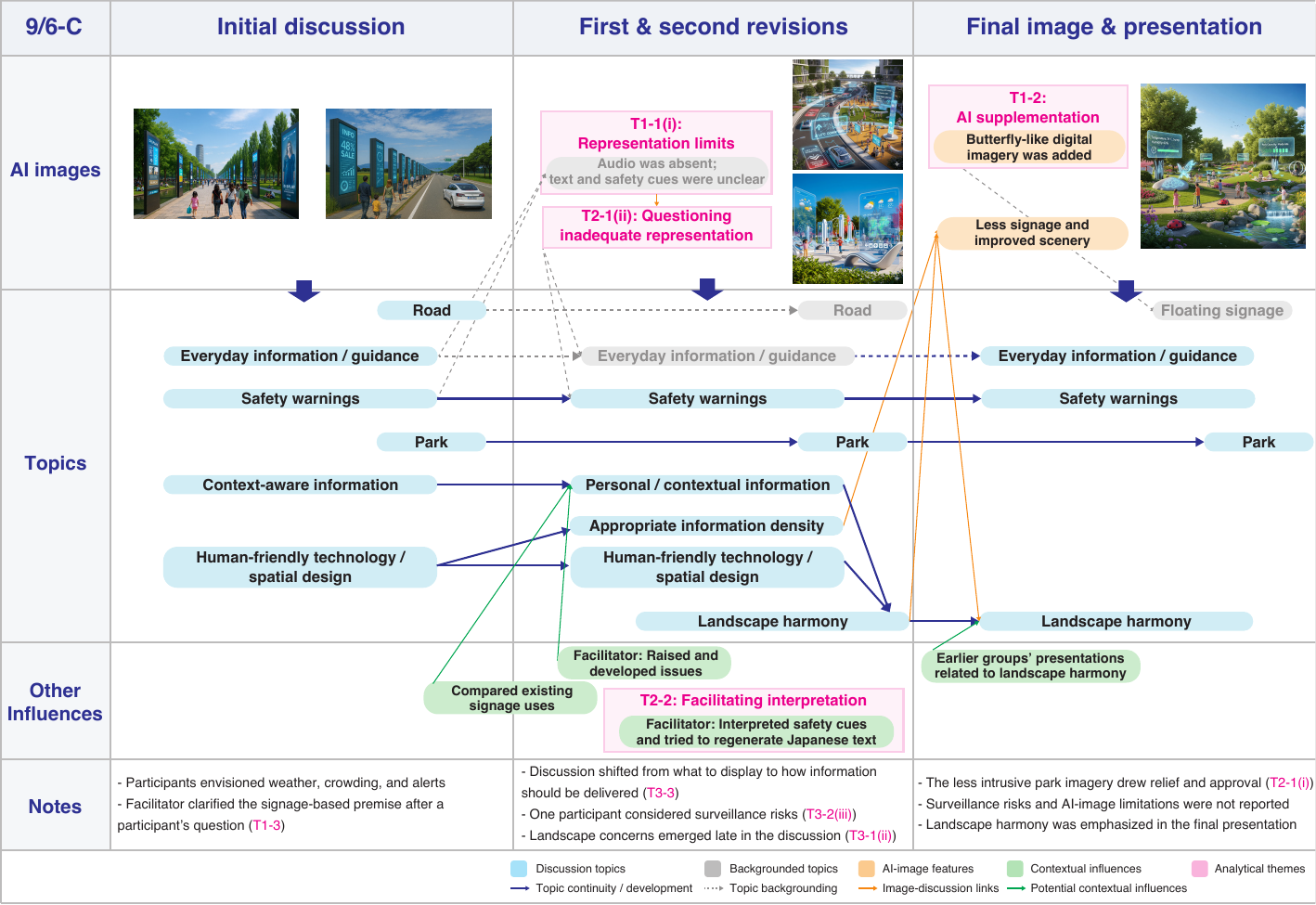}
  \caption{Evolution of topics and concerns in Group 9/6-C. Arrow and color conventions follow those in Figure~\ref{fig:results-3-chart0906A}.}
  \label{fig:results-3-chart0906C}
\end{figure}

Group 9/6-C initially discussed everyday information and safety warnings for roads and parks, including weather, crowding, and accident prevention (Figure~\ref{fig:results-3-chart0906C}). The facilitator also clarified the workshop's signage-centered premise in response to a participant's question (Theme 1-3). In the first two image revisions, audio was absent, and Japanese text and safety cues were unclear (Theme 1-1(i)). Participants questioned these inadequate representations (Theme 2-1(ii)), while the facilitator interpreted possible safety cues and attempted to regenerate Japanese text (Theme 2-2). Discussion then shifted from what information to display to how it should be delivered. Participants continuously developed ideas about personalized and context-aware information and appropriate information density (Theme 3-3). These exchanges drew on existing signage practices, while concerns about surveillance risks were raised but received limited further exploration (Theme 3-2(iii)). Toward the end, landscape harmony gained prominence (Theme 3-1(ii)), potentially informed by earlier groups' presentations (Theme 4-3). The final image depicted fewer displays and improved park scenery, alongside AI-added butterfly-like digital imagery (Theme 1-2). Participants welcomed the less intrusive environment (Theme 2-1(i)). The final presentation emphasized everyday guidance, safety, and landscape harmony, while personalized information, surveillance risks, and the limitations of AI-generated images were not fully conveyed.

\newpage
\subsubsection*{Group 9/6-D}
\begin{figure}[htb]
  \centering
  \includegraphics[keepaspectratio,width=1.0\linewidth]{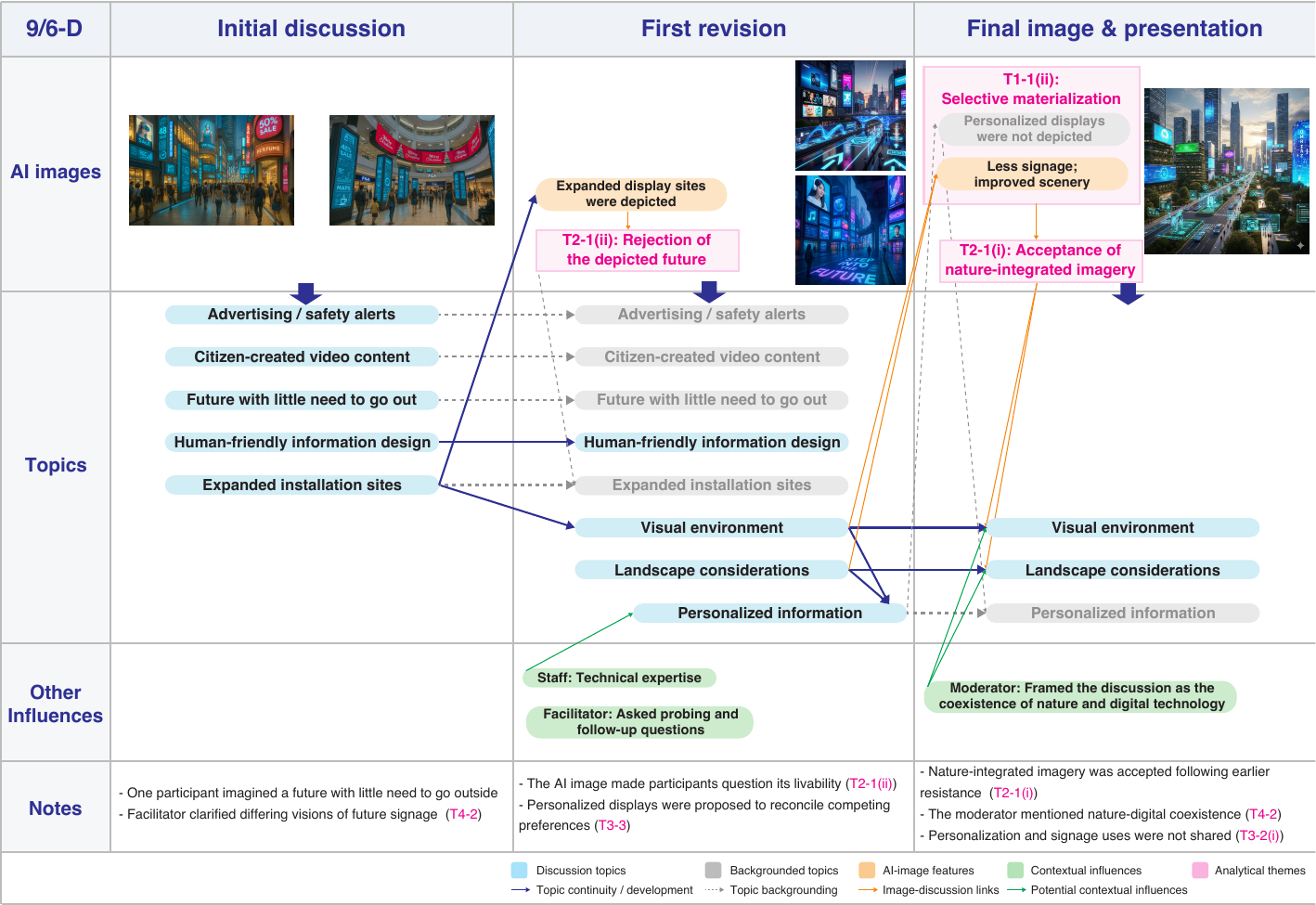}
  \caption{Evolution of topics and concerns in Group 9/6-D. Arrow and color conventions follow those in Figure~\ref{fig:results-3-chart0906A}.}
  \label{fig:results-3-chart0906D}
\end{figure}
Group 9/6-D initially proposed diverse ideas, including advertising and safety alerts, citizen-created video content, and expanded installation sites for digital signage (Figure~\ref{fig:results-3-chart0906D}). One participant also envisioned a future with little need to go outside, while the facilitator helped clarify differing visions of future signage (Theme 4-2). The first revised images depicted expanded display sites across urban infrastructure, but participants questioned the livability of the resulting highly digitized cityscape (Theme 2-1(ii)). Drawing on their experiences and preferences (Theme 4-3), they expressed concerns about its artificial appearance and lack of natural surroundings. The facilitator's follow-up questions encouraged further exploration of the visual environment, landscape considerations, and alternative futures (Themes 3-1(ii) and 4-2). Participants also proposed personalized information displays to reconcile differing preferences for digital technology and nature (Theme 3-3). The final image depicted less signage and improved scenery, which participants accepted after their earlier resistance (Theme 2-1(i)). However, the proposed personalized displays were not represented (Theme 1-1(ii)). During the final presentation, the moderator and participants framed the discussion around the coexistence of nature and digital technology (Theme 4-2), while personalized information and earlier signage applications were not shared (Theme 3-2(i)).

\newpage
\subsubsection*{Group 9/7-A}
\begin{figure}[htb]
  \centering
  \includegraphics[keepaspectratio,width=1.0\linewidth]{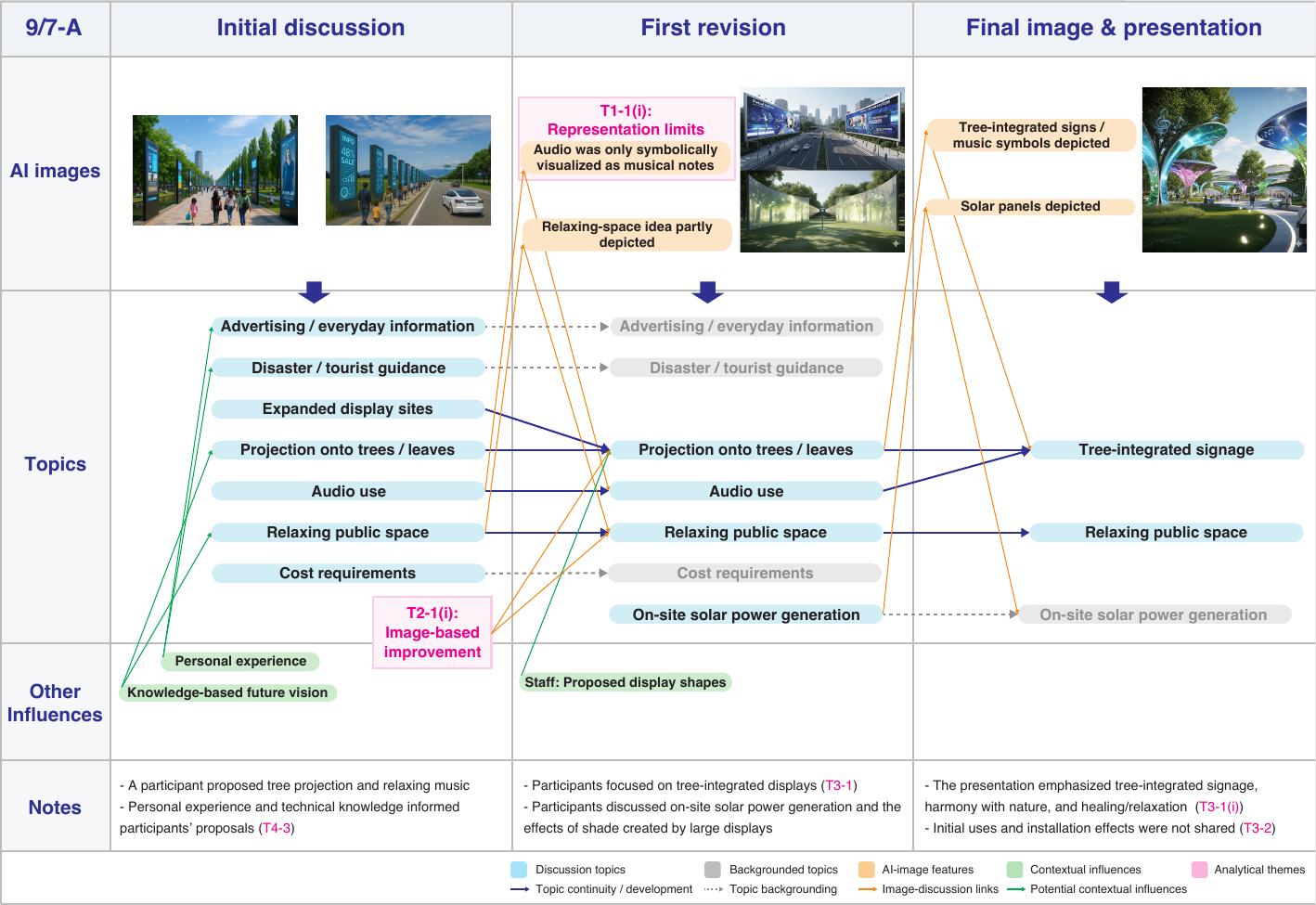}
  \caption{Evolution of topics and concerns in Group 9/7-A. Arrow and color conventions follow those in Figure~\ref{fig:results-3-chart0906A}.}
  \label{fig:results-3-chart0907A}
\end{figure}

Group 9/7-A initially proposed advertising and everyday information, disaster and tourist guidance, and expanded display sites, alongside projection onto trees or leaves and music to create relaxing public spaces (Figure~\ref{fig:results-3-chart0907A}). These ideas partially drew on participants' personal experiences and technical knowledge (Theme 4-3). The first revised images partly depicted the envisioned relaxing space, but audio was represented only symbolically through musical notes (Theme 1-1(i)). In response to these images, participants further developed tree-integrated signage, considering display shapes and installation methods with suggestions from workshop staff (Themes 2-1(i), 3-1(i), and 4-3). They also discussed the effects of shade from large displays and on-site solar power generation. Meanwhile, specific applications such as disaster and tourist guidance, as well as the intended uses of music, received less sustained attention (Theme 3-2). Nevertheless, the broader aim of creating relaxing public spaces persisted through discussions of landscape harmony (Theme 3-3). The final image depicted tree-integrated signage, music symbols, and solar panels. The final presentation emphasized harmony with nature and relaxation, while initial applications and practical considerations, including installation effects, were not shared (Theme 3-2).

\newpage
\subsubsection*{Group 9/7-B}
\begin{figure}[htb]
  \centering
  \includegraphics[keepaspectratio,width=1.0\linewidth]{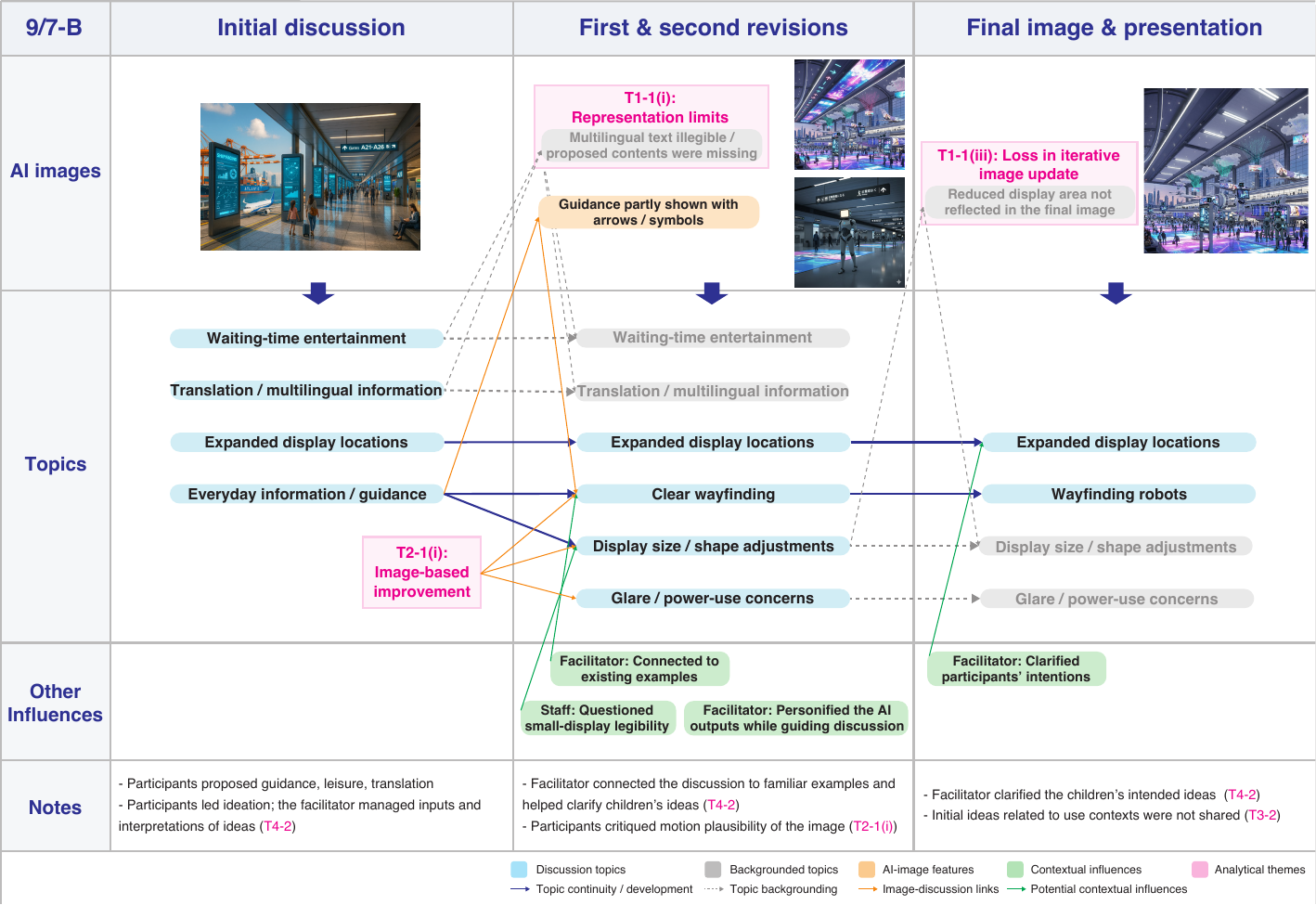}
  \caption{Evolution of topics and concerns in Group 9/7-B. Arrow and color conventions follow those in Figure~\ref{fig:results-3-chart0906A}.}
  \label{fig:results-3-chart0907B}
\end{figure}
Group 9/7-B initially proposed expanding digital signage to aircraft, ceilings, utility poles, and other locations, alongside everyday guidance, waiting-time entertainment, and multilingual information (Figure~\ref{fig:results-3-chart0907B}). Participants actively developed these ideas, proposing additional display locations such as clouds and floors, as well as robots (Theme 3-1(i)). In the first two image revisions, guidance was partly represented through arrows and symbols, but multilingual text was illegible and some proposed content was missing (Theme 1-1(i)). Participants critically examined the images, questioning the clarity of wayfinding cues and the plausibility of depicted movement (Theme 2-1(i)). They also proposed adjustments to display size and shape and raised practical concerns about glare and electricity consumption. Throughout these exchanges, the facilitator connected ideas to familiar examples, clarified the children's intentions, and occasionally personified the AI to manage the pace of discussion and image generation (Theme 4-2). Workshop staff also questioned whether smaller displays would remain legible. Although participants requested smaller displays, the final image did not clearly reflect a reduction in display area (Theme 1-1(iii)). The final presentation emphasized expanded display locations and wayfinding robots, while initial ideas about waiting-time entertainment and multilingual information, along with practical concerns about glare and power use, were not shared (Theme 3-2).

\newpage
\subsubsection*{Group 9/7-C}
\begin{figure}[htb]
  \centering
  \includegraphics[keepaspectratio,width=1.0\linewidth]{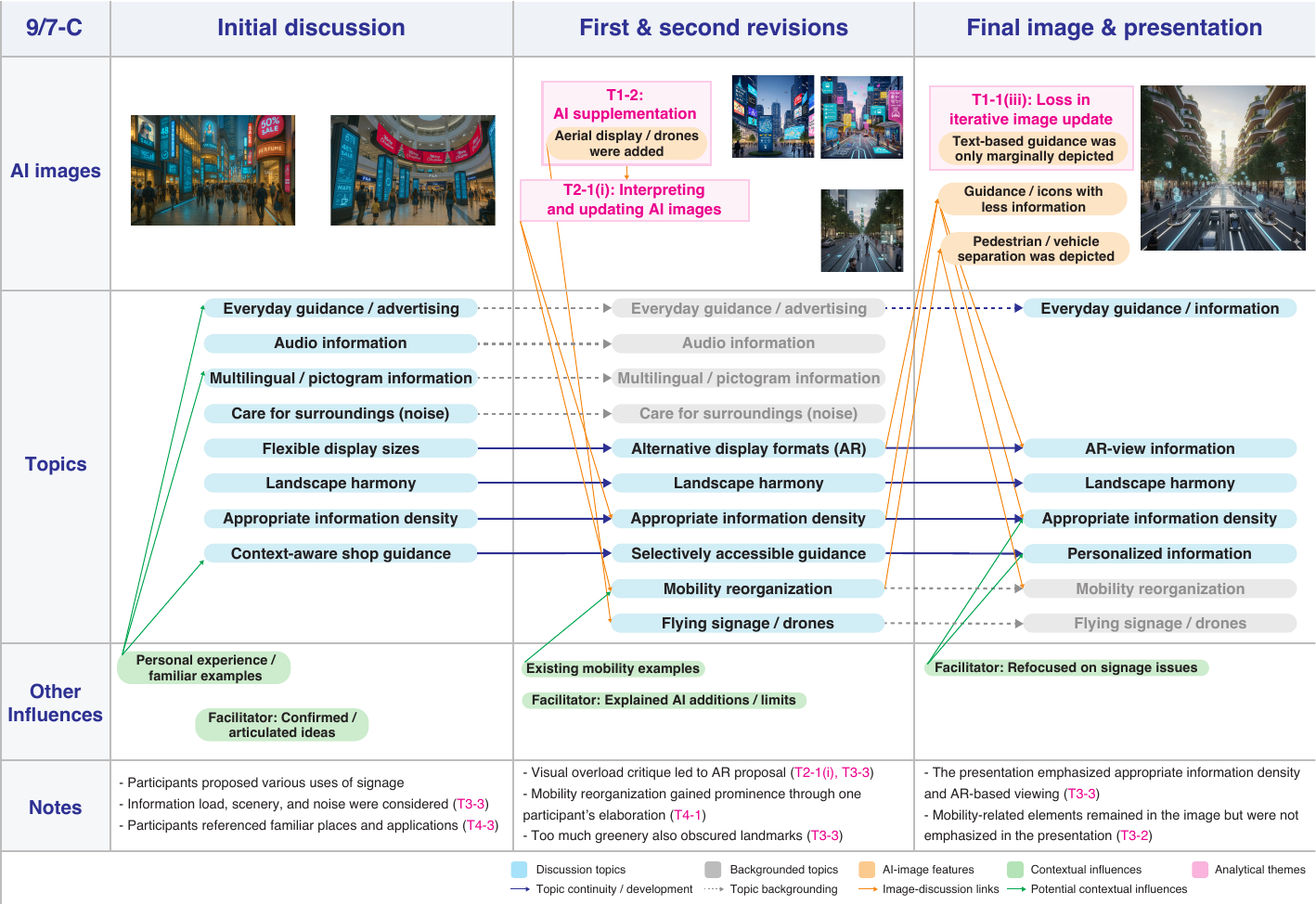}
  \caption{Evolution of topics and concerns in Group 9/7-C. Arrow and color conventions follow those in Figure~\ref{fig:results-3-chart0906A}.}
  \label{fig:results-3-chart0907C}
\end{figure}
Group 9/7-C initially proposed everyday guidance and advertising using audio, multilingual information, and pictograms, alongside context-aware shop guidance and flexible display sizes (Figure~\ref{fig:results-3-chart0907C}). Participants also emphasized appropriate information density and care for the surrounding environment, drawing on familiar places and personal experiences (Theme 4-3). In the first two image revisions, AI-generated images introduced unexpected elements, including aerial displays, drones, and butterflies (Theme 1-2). Participants noticed these additions without necessarily adopting them, and the facilitator helped distinguish AI-added elements from participants' proposals (Themes 2-1(i) and 2-2). Criticism of excessive visual information led to proposals for alternative display formats, particularly selectively accessible AR-based guidance, extending the initial concerns about information density and context-aware information (Theme 3-3). In parallel, one participant drew on existing transportation technologies to propose moving walkways, pedestrian–vehicle separation, and underground transportation, bringing mobility reorganization into focus (Theme 4-3). Participants also questioned an image in which increased greenery obscured familiar landmarks, maintaining their concern for landscape harmony (Theme 3-3). The final image depicted guidance icons with less information and pedestrian–vehicle separation, although text-based guidance was only marginally represented (Theme 1-1(iii)). At the final stage, the presentation emphasized appropriate information density and AR-based viewing (Theme 3-3). Mobility-related ideas remained partly visible in the image but were not emphasized in the presentation, while earlier audio and multilingual communication received limited further attention (Theme 3-2).

\newpage
\subsubsection*{Group 9/7-D}
\begin{figure}[htb]
  \centering
  \includegraphics[keepaspectratio,width=1.0\linewidth]{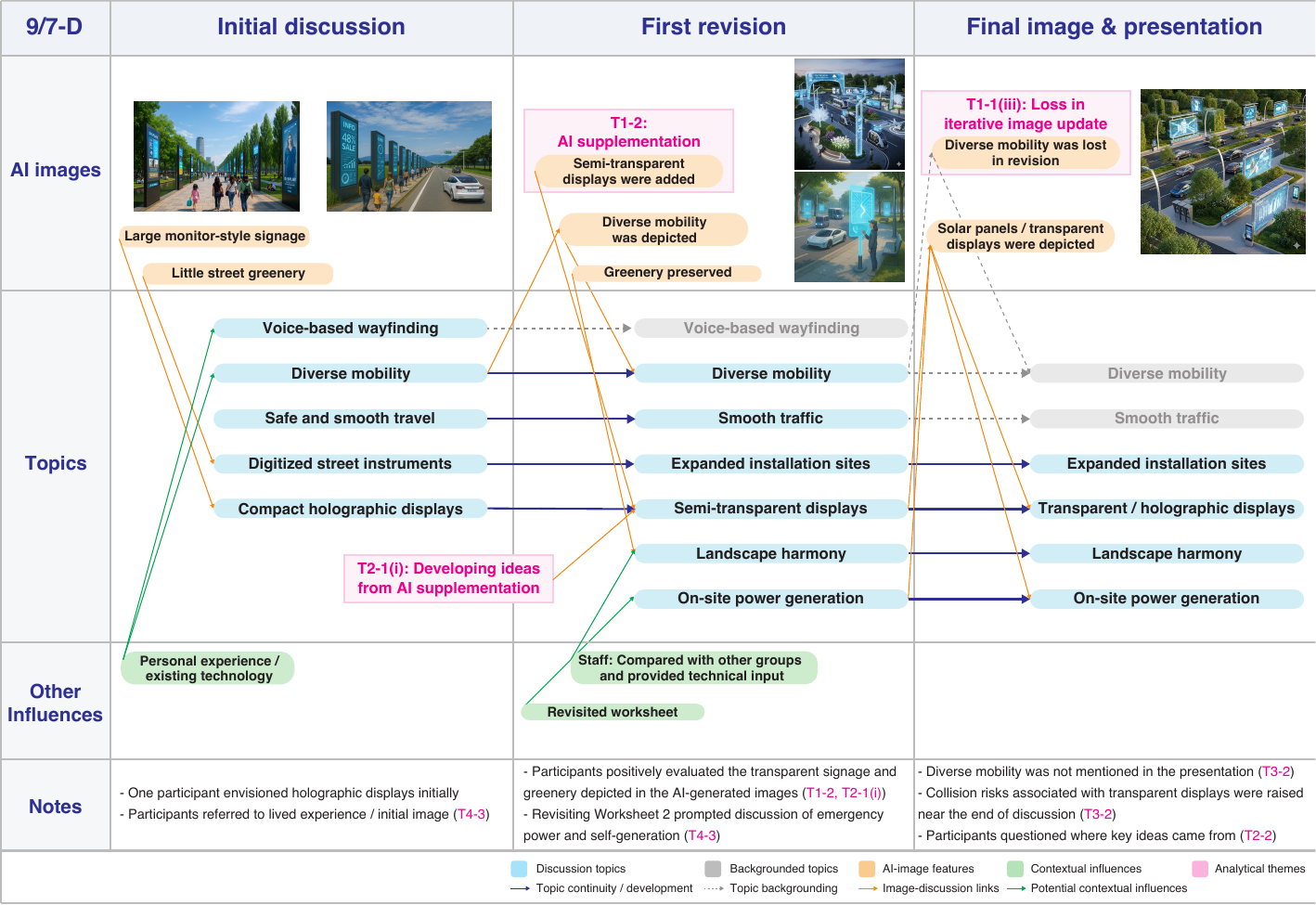}
  \caption{Evolution of topics and concerns in Group 9/7-D. Arrow and color conventions follow those in Figure~\ref{fig:results-3-chart0906A}.}
  \label{fig:results-3-chart0907D}
\end{figure}
Group 9/7-D initially discussed voice-based wayfinding, diverse mobility and safe travel, digitized street infrastructure, and compact holographic displays (Figure~\ref{fig:results-3-chart0907D}). The first revised images depicted diverse mobility and preserved greenery, while introducing semi-transparent signage that participants had not explicitly requested (Theme 1-2). Participants welcomed this addition and developed it into a proposal for displaying information without disrupting the cityscape, emphasizing landscape harmony (Theme 2-1(i)). Discussion also addressed practical requirements. Revisiting an earlier worksheet prompted consideration of emergency power supply and on-site power generation, while a workshop staff member explained how unconventional display shapes could be realized using projectors and flexible screens (Theme 4-3). In a subsequent image, the diversity of mobility options was no longer evident (Theme 1-1(iii)), although semi-transparent displays and solar panels remained visible. Participants also raised potential collision risks associated with semi-transparent displays, but these received limited further exploration (Theme 3-2(iii)). Toward the end, participants reflected on how AI-generated additions might have shaped their collective vision (Theme 2-2). The final presentation emphasized compact and semi-transparent displays, landscape harmony, and emergency power, while earlier ideas about diverse mobility and voice-based guidance were not shared (Theme 3-2).



\end{document}